\documentclass[11pt]{article}

\usepackage[utf8]{inputenc}
\usepackage[T1]{fontenc}
\usepackage{lmodern}
\usepackage[a4paper,
            right=3cm,
            top=2.5cm,
            bottom=2.5cm]{geometry}
\usepackage{amsmath, amssymb, amsfonts} 
\usepackage{graphicx} 
\usepackage{caption}
\usepackage{subcaption}
\usepackage{booktabs} 
\usepackage{hyperref} 
\usepackage{float} 
\usepackage{xcolor}
\usepackage{verbatim}
\usepackage{booktabs}
\usepackage{makecell}
\usepackage{csquotes}
\usepackage{authblk}
\usepackage{pdflscape}
\usepackage[
  backend=biber,
  style=apa,
  doi=true,
  isbn=false,
  url=true,
  giveninits=true,
  uniquename=false,
  hyperref=true,
  sorting=nyt,
]{biblatex}
\usepackage{fvextra}
\usepackage{etoolbox}

\DefineVerbatimEnvironment{CompactVerbatim}{Verbatim}{
  breaklines=true,
  breakanywhere=true,
  breaksymbolleft={},
  breaksymbolright={},
  fontsize=\small,
  baselinestretch=0.95,
  listparameters={
    \setlength{\topsep}{2pt}
    \setlength{\partopsep}{0pt}
    \setlength{\parsep}{0pt}
    \setlength{\itemsep}{0pt}
  }
}
\usepackage[bottom]{footmisc}
\newtheorem{theorem}{Theorem}[section]
\newtheorem{definition}[theorem]{Definition}

\BeforeBeginEnvironment{CompactVerbatim}{\vspace{4pt}}
\AfterEndEnvironment{CompactVerbatim}{\vspace{4pt}}

\newcounter{compactverbatim}

\title{Cone-Based Statistics for Characterizing the  Local Morphology of High-Dimensional Point Clouds}

\author[1,2]{Lynn Delcon\thanks{Corresponding author \href{mailto:Lynn.Delcon@vub.be}{Lynn.Delcon@vub.be}}}
\author[3]{David Tewodrose}
\author[1,2]{Andres Algaba}
\author[1,2,4]{Vincent Ginis}

\affil[1]{%
    Department of Business Technology and Operations, Data Analytics Laboratory, Vrije Universiteit Brussel,
    Pleinlaan 2, 1050 Brussels, Belgium}
\affil[2]{imec-SMIT, Vrije Universiteit Brussel,
Pleinlaan 9, 1050 Brussels, Belgium}
\affil[3]{Department of Mathematics and Data Science, Vrije Universiteit Brussel, 1050 Brussels, Belgium}
\affil[4]{School of Engineering and Applied Sciences, Harvard University,
Cambridge, Massachusetts 02138, USA}

\date{\today}

\begin{document}

\maketitle

\begin{abstract}
  
We introduce a cone-based local geometric statistic\footnote{Code publicly available on \href{https://github.com/lynndelcon/A-Cone-Based-Local-Geometric-Statistic-for-Identifying-Candidate-Boundary-Points-and-their-Concavity}{GitHub}.} for characterizing the angular configuration of candidate boundary points in high-dimensional point clouds and investigate its impact on cosine-similarity-based metrics when reducing embedding dimensions.
Our boundary score, $BoS(x)\in[0,1]$, measures the proportion of projected neighbors contained in a cone defined by two maximally separated neighbor directions. Using the cone aperture expressed in degrees, each point is assigned a convexity behavior. We derive two variants of our cone-based statistic to account for concavity and for a boundary-relative centrality measure, which we refer to as oriented depth. The benchmark evaluations on synthetic datasets show comparable AUROC values of our statistics compared to state-of-the-art methods in two boundary settings: the distance to the support setting and the position of a point with respect to its neighbors, which we refer to as the local-sector setting. We observe the added-value of our second variant to better detect and characterize local concavity in $2$ dimensions compared to the other two cone-based statistics. 
All three cone-based statistics provide us with different but complementary frameworks to characterize the morphology of high-dimensional point clouds.
Among the tested boundary-guided PCA approaches, scaling observations with their persistent-boundary scores before computing the covariance matrix generally yields smaller deviations from the original embedding space in terms of mean positive–negative cosine-similarity margins and pairwise win rates. These application results are relevant for optimizing RAG workflows with respect to storage resources.
\end{abstract}


\section{Introduction}\label{sec:related_work_bpdm}

Research on boundary points of point clouds in $\mathbb{R}^D$ has developed mainly along two lines: clustering (\cite{DBLP}; \cite{CLG}; \cite{DCM}; \cite{3_way}; \cite{LoDD}) and image reconstruction (\cite{angular_gap}; \cite{mineo2019novel}). The majority of boundary point detection methods in clustering analysis are density-based, whereas those applied in image reconstruction are geometry-based. The former methods consider boundary points as low-density points; the neighborhood of such points is widespread. An interior point is then typically located in a high density region; its neighbors are very close. In contrast, geometry-based methods use other metrics to identify boundary points such as the tangent plane (\cite{tangent}), the angular gap (\cite{angular_gap}), the hyperplane or the Directional Uniformity (DU) (\cite{hickok2026boundarydetectionhighdimensionalpoint}). These metrics are still provided by the neighborhood configuration.

Both types of methods allow us to distinguish border points from edge points. A border point is typically an intermediate point that constitutes a bridge between two clusters, whereas an edge point is rather on the external contours of the point cloud (\cite{tangent}; \cite{3_way}). While the density-based algorithm of \textcite{3_way} uses a classification-overlap concept to distinguish border points from edge ones, the tangent plane method of \textcite{tangent} suggests varying the neighborhood size, i.e., the number of $k$ nearest neighbors ($k$-nn). Points detected as boundary ones for a low neighborhood size (say $10$-nn) and for a medium-to-high neighborhood size (say $30$-nn) are considered as edge points. Border points are then defined as transient boundary points.

Although the geometry-based methods of \textcite{tangent} and \textcite{hickok2026boundarydetectionhighdimensionalpoint} are high-performing boundary point detectors, the method we propose provides us with the additional local angular configuration of candidate boundary points.
Our method enters the geometry-based category  and is applicable in high dimensions. It uses the proportion of nearest neighbors as in \textcite{tangent} to compute the boundary score, and the angular gap concept from \textcite{angular_gap} to capture pointwise concavity. 
Throughout this work, convexity and concavity refer to the angular configuration of neighboring points in a selected projection plane, rather than to differential-geometric curvature of the underlying support.
Lastly, distinguishing border from edge points can be realized using the neighborhood resolution concept of \textcite{tangent}.

\paragraph{Structure.} In the remainder, we present our cone-based boundary point detection method and two variants. We then benchmark all three with respect to state-of-the-art methods on two types of boundary point settings. We conclude with the application of boundary scores to reduce the embedding dimensions and investigate their impacts on query–document cosine-similarity margins relative to the original embedding space.

\section{Proposed Boundary Point Detection Method}\label{sec:our_method}

Let us provide a mathematical description of our method. For $u, v \in \mathbb{R}^D \backslash \{0\}$, let
\[
\theta(u,v) = \frac{180^\circ}{\pi} \arccos \left( \frac{<u,v>}{||u|| \ ||v||} \right)
\]
be the measure in degree of the angle between $u$ and $v$. Here, $<,>$ is the Euclidean scalar product and $||\cdot ||$ is the associated norm. The convex cone generated by $u$ and $v$ is
\[
\mathcal{C}(u,v) = \{su + tv : s,t \ge 0\}.
\]

Consider a point cloud $\mathcal{X} \subset \mathbb{R}^D$, a positive integer $K \ge 3$, and a given point $x \in \mathcal{X}$. Denote by $\{x_j\}$ the $K$ nearest neighbors of $x$, where the distance between points in $\mathcal{X}$ is the Euclidean one. For any $j \in \{1,\ldots,K\}$, set
\[
v_{j}=x_j-x.
\]

\subsection*{Initial method}\label{var1}

Reorder the vectors $v_1,\ldots,v_K$ in such a way that
\[
(v_1,v_2) \in \underset{1^\circ\leq\theta(v_j,v_l)\leq179.9^\circ}{\arg\max} \theta(v_j,v_l).
\]
The constraint on $\theta(v_j,v_l)$ is meant to exclude co-linear vectors. Set $J = \{3,\ldots,K\}$.  Consider the orthogonal projections $\{y_j\}_{j \in J}$ onto $P = \mathrm{span}(v_{1},v_{2})$ of the remaining vectors $\{v_j\}_{j \in J}$. To determine $N(x) = \# \{ j \in J : y_j \in \mathcal{C}(v_{1},v_{2})\}$, compute the sum $\theta(v_{1},y_j) + \theta(y_j,v_{2})$ which, for $y_j \neq 0$, is equal to $\theta(v_{1},v_{2})$\footnote{In the code, we use an angular comparison tolerance of $0.0001$.} if and only if $y_j \in \mathcal{C}(v_{1},v_{2})$. Call the resulting proportion $BoS(x)=N(x)/(K-2)$ the boundary score of $x$. For the edge case where $y_j =0$, this neighbor is not considered as inside the cone but still counts in the denominator, lowering $BoS(x)$. 

\begin{figure}[H]
    \centering
    \includegraphics[width=0.4\linewidth]{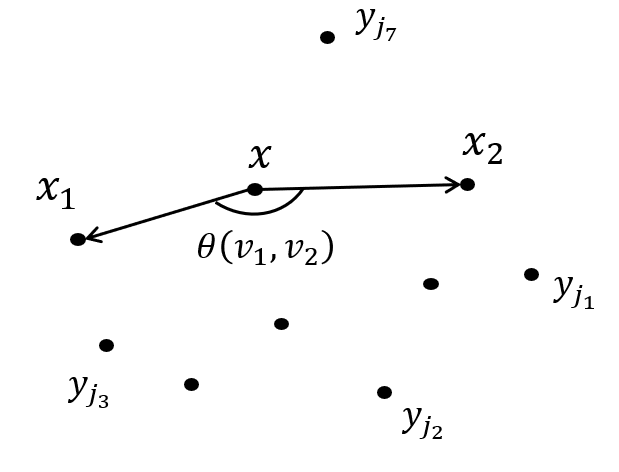}
    \caption{Illustration of a case where $K=9$ and $BoS(x) = 6/7 \approx 0.86$.}
    \label{fig:initial_sketch}
\end{figure}

Let $B=\partial_P\mathcal{C}(v_1,v_2)$ denote the relative boundary of the cone in $P$, consisting of the two rays generated by $v_1$ and $v_2$. A boundary score near $1$ means that most of the projected neighbors lie within the selected cone, suggesting a candidate convex angular configuration in the projection plane.
Conversely, a boundary score near $0$ indicates that most of the projected neighbors lie outside the selected cone, suggesting a candidate concave angular configuration in the projection plane. This leads us to the following definition.

\begin{definition}
We designate $x$ as a candidate boundary point when $$BoS(x)\geq BoS\_THRESH \quad \text{or} \quad BoS(x)\leq 1-BoS\_THRESH$$where $BoS\_THRESH$ is a sufficiently high hyperparameter. We label $x$ as a convex boundary point in the former case and a concave one in the latter.
\end{definition}

The value of BoS\_THRESH determines the set of points labeled as boundary points.

\subsection*{Variant 1: Adjusted Angles}\label{var1}

In order to measure concavity at boundary points, we propose a refinement of the initial method which we call Adjusted Angles. It consists of opening the cone $\mathcal{C}(v_1,v_2)$ so that all the projections $\{y_j\}$ lie either inside or outside a new cone. The aperture of this adjusted cone measures concavity, which provides an additional characterization of the local angular geometry. This refinement uses two extra hyperparameters:
\begin{itemize}
\item the proportion threshold PROP\_THRESH $\in (0.5,1)$ singles out points where the concavity is evaluated. Setting this threshold to $0.51$ increases the amount of candidate concave boundary points, while a higher threshold decreases the exploration rate.
\item the angular gap ANGULAR\_GAP $ \in (0^\circ,180^\circ]$ defines the empty sector opening in the neighborhood of $x$ such that a medium-to-high angular gap implies that $x$ is at the boundary of its neighborhood. An angular gap of $0^{\circ}$ corresponds to an interior point. \textcite{angular_gap} suggest to set this hyperparameter to $90^{\circ}.$
\end{itemize}

Scenario 1. $BoS(x) \ge \text{PROP\_THRESH}$.

Consider $J_{\text{out}} = \{j \in J :  y_j \notin \mathcal{C}(v_1,v_2) \}$. For $\ell \in \{1,2\}$, define $\alpha_{j,\ell} = \theta(y_j,v_\ell)$ for any $j \in J_{\text{out}}$, then set $\beta_\ell^* = \max_{j \in J_{\text{out},\ell}}\alpha_{j,\ell}$ where $J_{\text{out},\ell} = \{j \in J_{\text{out}} : \min (\alpha_{j,1},\alpha_{j,2})=\alpha_{j,\ell}\}$. Lastly, set
$$\omega^{\text{high BoS}}:=\theta(v_1,v_2) +\beta^{*}_{1}+ \beta^{*}_{2}.$$
Case 1. $\omega^{\text{high BoS}} \geq 180^{\circ}$. If $360^{\circ}-\omega^{\text{high BoS}} \geq$ ANGULAR\_GAP, the adjusted angle becomes $360^{\circ}-\omega^{\text{high BoS}}$ and the associated $BoS(x)$ is $0$; all projected neighbors are outside the new cone of aperture $360^{\circ}-\omega^{\text{high BoS}}$. In this case, we say that the cone is concave.\\
Case 2. $\omega^{\text{high BoS}} < 180^{\circ}$. No angular constraint needs to be met given that the empty sector is the complement of $\omega^{\text{high BoS}}$. Hence, $\omega^{\text{high BoS}}$ defines the adjusted angle and the associated $BoS(x)$ is $1$; all projected neighbors are inside the enlarged convex cone.\\

\begin{figure}[H]
    \centering
    \includegraphics[width=0.8\linewidth]{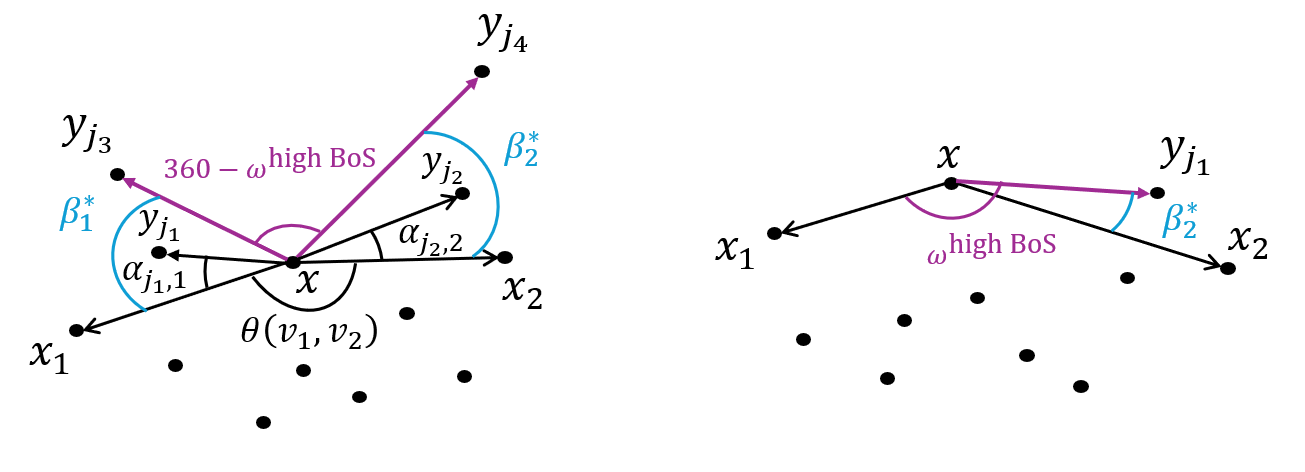}
    \caption{Illustration for $BoS(x) \geq \text{PROP\_THRESH}$. Recall that $v_1 = x_1 -x$, $v_2=x_2 - x$ and  $j\in J_{\text{out}}$. Case 1 (left), Case 2 (right). In both cases, we evaluate concavity in $P$, the span of $v_1,v_2$.}
    \label{fig:adjusted_sketch}
\end{figure}

Scenario 2. $BoS(x) \le 1- \text{PROP\_THRESH}$.

Consider $J_{\text{in}} = \{j \in J : y_j \in \mathcal{C}(v_1,v_2)\}$. 
For $\ell \in \{1,2\}$, define $\alpha_{j,\ell} = \theta(y_j,v_\ell)$ for any $j \in J_{\text{in}}$, then set $\beta_\ell^* = \max_{j \in J_{\text{in},\ell}}\alpha_{j,\ell}$ where $J_{\text{in},\ell} = \{j \in J_{\text{in}} : \min (\alpha_{j,1},\alpha_{j,2})=\alpha_{j,\ell}\}$. Lastly, set
$$\omega^{\text{low BoS}}:=\theta(v_1,v_2) -\beta^{*}_{1}- \beta^{*}_{2}.$$

If $\omega^{\text{low BoS}}\geq$ANGULAR\_GAP, the adjusted angle is defined as $\omega^{\text{low BoS}}$ and the associated $BoS(x)$ is $0$; the cone is shrunk inwards and all projected neighbors are outside of it. In this case, we also say that the cone is concave.\\

\begin{figure}[H]
    \centering
    \includegraphics[width=0.35\linewidth]{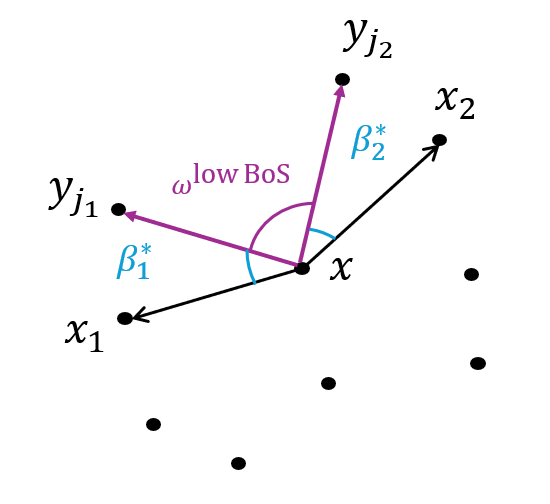}
    \caption{Illustration for $BoS(x) \leq \text{1-PROP\_THRESH}$, $j\in J_{\text{in}}$.}
    \label{fig:adjusted_sketch}
\end{figure}
In both scenarios, if the angular constraint is not met, the adjusted score and angle fall back to the initial ones.

\subsection*{Variant 2: Oriented Angles}

The oriented-angle variant constrains the selection of neighboring directions using their alignment with the direction towards the point-cloud barycenter. Let $g$ be the barycenter of $\mathcal{X}$. We define the set $VAL$ of couples $(v_j,v_l)$ satisfying the alignment constraint
\begin{equation*}
\frac{<v_j + v_l,g -x>}{||v_j+v_l|| \, ||g -x||} \geq \text{DIR\_THRESH}
\end{equation*}
where the direction threshold DIR\_THRESH $\in (0,1)$ is a hyperparameter. Then we define
\begin{equation*}
(v_1,v_2) = \underset{1^{\circ}\leq(v_j,v_l)\in VAL\leq 179.9^{\circ}}{\arg\max} \theta(v_j,v_l), \qquad \theta_{\text{oriented}} = \theta(v_1,v_2).
\end{equation*}

We compute the oriented boundary score $BoS(x)_{\text{oriented}}$ with $\mathcal{C}(v_1,v_2)$ as in the initial method. The higher the parameter DIR\_THRESH  is, the more difficult it may be to obtain the oriented angle $\theta_{\text{oriented}}$.

\begin{figure}[H]
    \centering
    \includegraphics[width=0.45\linewidth]{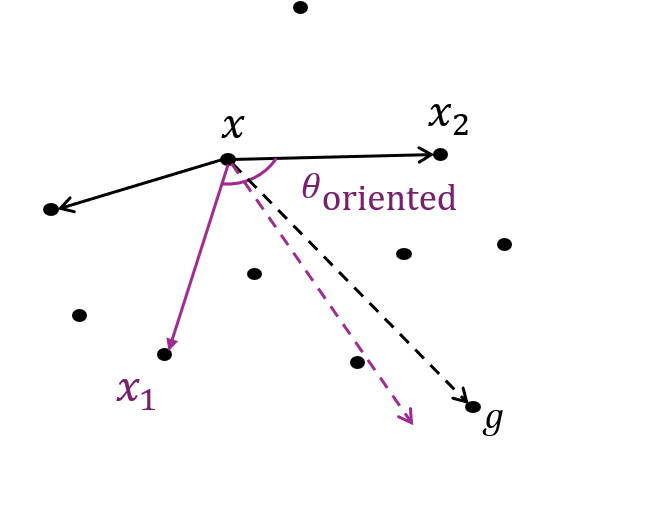}
    \caption{Illustration of the computation of $BoS(x)$ under the alignment constraint.}
    \label{fig:oriented_sketch}
\end{figure}

We underline two limitations of this variant. First, the alignment constraint towards the barycenter is dependent of the sufficient number of neighbors $k$ and data points $n$ with respect to the dimension; the increase of dimensions implies a decline in density if the number of observations does not increase accordingly (curse of dimensionality). Second, the barycenter is sensitive to the distribution of the points. While this method is well suited to convex supports with an appropriately positioned barycenter, it can also perform well on certain non-convex supports, such as star-shaped domains. Its reliability depends on the positioning of the barycenter relative to the support geometry, particularly when the point-cloud barycenter differs from that of the underlying support \(\Omega\).

To conclude, the oriented boundary score induced by this method can be considered as a boundary-relative centrality measure that we refer to as oriented depth. A $BoS(x)_{\text{oriented}}$ near $0.5$ reflects points located near the barycenter, hence, the farthest from the border. A $BoS(x)_{\text{oriented}}$ near $1$ represents points on the support boundary. Therefore, we define the depth as follows,

\begin{equation*}
\text{depth}_{\text{oriented}}(x) = 1-2|BoS(x)_{\text{oriented}}-0.5| \in [0,1].
\end{equation*}

\subsection{Simulation in $\mathbb{R}^2$ and $\mathbb{R}^3$}\label{sec:sim}

\begin{figure}[H]
\centering
\includegraphics[width=1\linewidth]{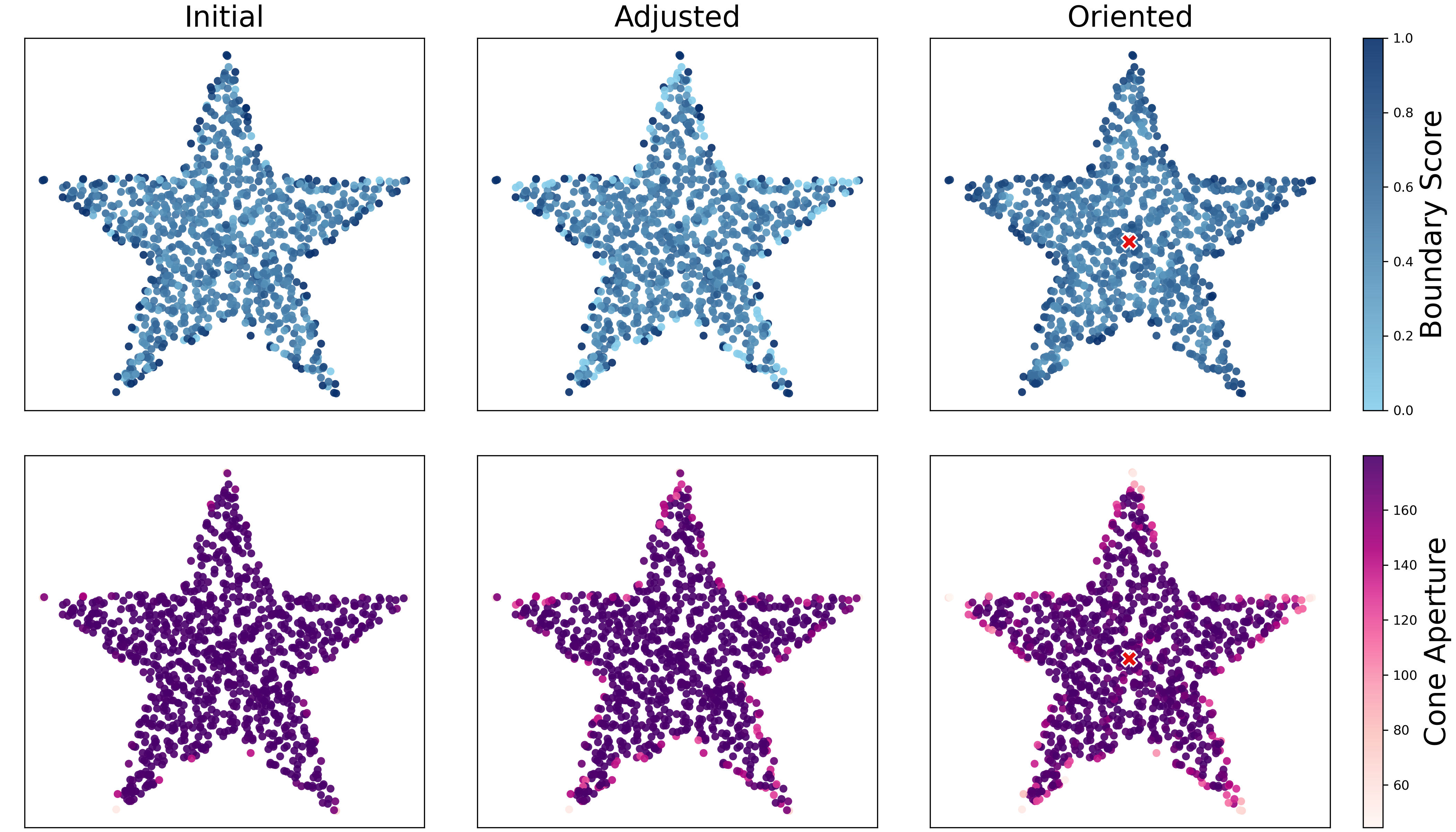}
\caption{Illustration of our methods in the star dataset. }
\label{fig:star_22nn}
\end{figure}

\begin{figure}[H]
\centering
\includegraphics[width=0.65\linewidth]{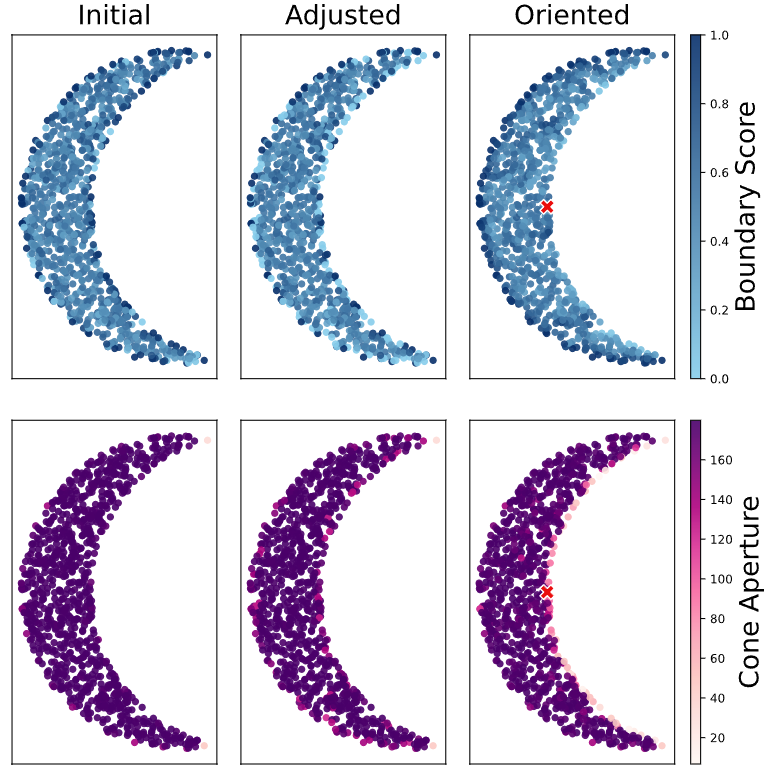}
\caption{Illustration of our methods in the moon dataset.}
\label{fig:moon_22nn}
\end{figure}

\begin{figure}[H]
\centering
\includegraphics[width=1\linewidth]{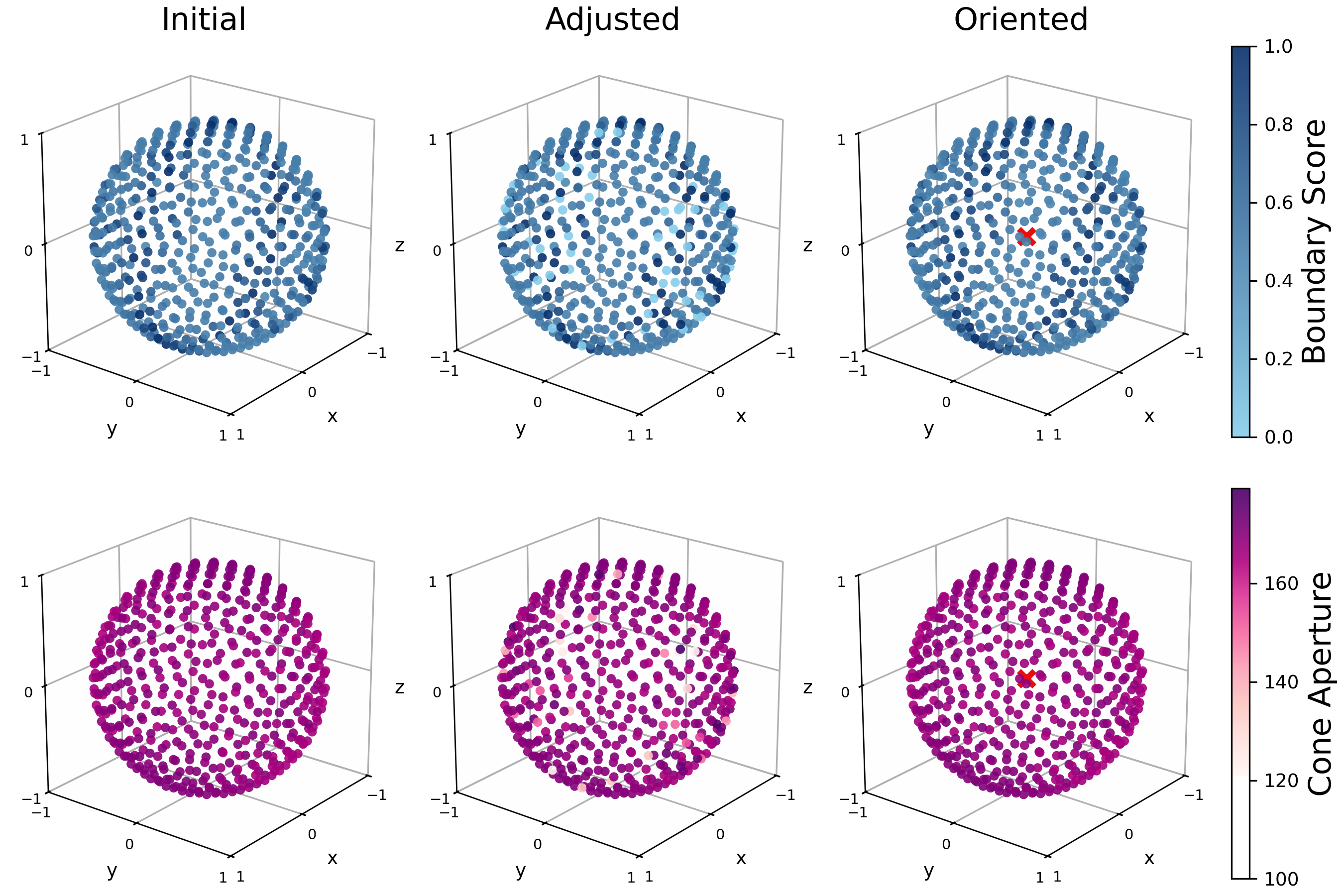}
\caption{Illustration of our methods in the Fibonacci Sphere dataset.}
\label{fig:fibo_22nn}
\end{figure}

We report the results on three toy datasets, namely, the star dataset ($n=1000$) (Figure \ref{fig:star_22nn}), the moon dataset ($n=1000$) (Figure \ref{fig:moon_22nn}) and the Fibonacci sphere dataset ($n=500$) (Figure \ref{fig:fibo_22nn}). For each set, we compute all three variants of the boundary score and angle for a neighborhood of $22-nn$. We set PROP\_THRESH $=0.51$ and ANGULAR\_GAP $=120^{\circ}$ for the adjusted method, and  DIR\_THRESH $=0.7$ for the oriented one. For the latter variant, we represent the barycenter with a red cross.

These simulations illustrate the qualitative behavior of all three variants. We highlight the changes brought by the adjusted variant compared to the initial method that provides us with smaller angles (adjusted cone aperture) and more boundary points.

The illustration in $\mathbb{R}^{3}$ shows us the variability of the boundary scores among similar data points. This property is explained by the increase of possible cone inclinations in higher dimensions. Depending on the direction of the cone formed by the two basis vectors, few neighbors can be projected outside of the cone and hence push or pull $BoS(x)$ towards $0.5$. This motivates the use of the adjusted variant to detect more boundary points, setting PROP\_THRESH < BoS\_THRESH. Concerning the oriented variant, the observation in the moon dataset of very small angles (cone opening) comes from the alignment constraint towards the barycenter $g$ (red cross) that implies a restricted projection area for the neighbors and hence a small $BoS(x)_{\text{oriented}}$.

Finally, we would like to emphasize that the boundary and concavity behaviors provided by our method reflect the local configuration of the surrounding neighbors and not the global point-cloud shape.

\section{Comparative Analyses}\label{sec:benchmark}

We test the performance of our cone-based methods against the $3$ SOTA methods discussed in Section \ref{sec:related_work_bpdm}, namely, the tangent plane (\cite{tangent}), the DU and the Hyperplane methods  (\cite{hickok2026boundarydetectionhighdimensionalpoint}).

In the construction of our benchmark datasets, we distinguish two boundary point categories: the \textit{support} boundary points and the \textit{local-sector} boundary points.

\paragraph{Support Boundary.} The first setting uses the Euclidean distance to the analytical boundary support, denoted as $dist(x,\partial \Omega)$,

\begin{equation*}
  dist(x,\partial \Omega)
  =\inf_{y\in\partial\Omega}\|x-y\|_2.
  \label{eq:support-distance-general}
\end{equation*}

The binary labeling of a point being a boundary or an interior point is respectively defined as follows,

\begin{equation*}
\qquad
y_i=\begin{cases}1&dist_i\leq\mathcal{Q}_{0.20}(dist),\\
0 &dist_i\geq\mathcal{Q}_{0.50}(dist),\\
&\text{excluded otherwise}.
\end{cases}
\end{equation*}
where $y_i$ is the boundary label assigned to $x_i\in \mathcal{X}$, $dist_i$ refers to $dist(x_i,\partial \Omega)$, and $\mathcal{Q}_{0.20}(dist)$ and $\mathcal{Q}_{0.50}(dist)$ denote the $20^{th}$ and $50^{th}$ percentile, respectively, of the empirical distance distribution. This percentile-based rule compares boundary points to deep-interior points and allows for both classes to contain enough observations. Inspired by \textcite{hickok2026boundarydetectionhighdimensionalpoint}, intermediate points, that are neither boundary nor deep-interior points, are excluded from the AUROC computation. 

\paragraph{Local-sector Boundary.}
The second setting defines boundary points as points whose neighborhoods contain an empty angular region  (or an empty sector). We construct the local sector in the intrinsic space $\mathbb{R}^d$ by restricting neighboring directions to a spherical cap centered on a unit axis $u\in S^{d-1}$. Neighbors are generated as $x_j=x_i+r_jv_j$, where the unit directions $v_j$ are sampled uniformly with respect to the spherical surface measure on the cap $\mathcal{C}_d(u,\alpha)$ where $\alpha$ is the full angular opening of the occupied region,

\[
\mathcal{C}_d(u,\alpha)=\{v\in S^{d-1}:v^\top u\geq\cos(\alpha/2)\},
\]

and the radial distances $r_j$ are independently sampled from $\mathcal{U}(0.2,0.9)$.
For $d=2$, directions lie on the unit circle $S^1$; for $d=3$, they lie on the sphere $S^2$; and for $d=5$, on the four-dimensional sphere $S^4$. We define the missing fraction $f_d(x_i)$ as the proportion of the intrinsic directional space $S^{d-1}$ outside the chosen spherical cap,
\begin{align*}
    f_d(x_i) &= 1-
    \frac{\int_0^{\alpha/2}\sin^{d-2}(t)\,dt}
    {\int_0^\pi\sin^{d-2}(t)\,dt},\\
    y_i &=
    \begin{cases}
        1 & \text{if } f_d(x_i)\geq\tau,\\
        0 & \text{otherwise}.
    \end{cases}
\end{align*}
As we use isometric embedding in our dataset to increase the ambient dimension $D$, the intrinsic angular geometry and missing fraction are preserved (Annex \ref{annex:support-dataset}).

For $d=2$, the threshold $\tau=0.5$ corresponds to an empty angular sector of at least $180^\circ$ and favors methods using half-plane logic such as the tangent plane (\cite{tangent}), the hyperplance and the DU methods (\cite{hickok2026boundarydetectionhighdimensionalpoint}). We instead choose $\tau=1/3$, corresponding to an empty sector of at least $120^\circ$ for $d=2$, consistent with the choice of ANGULAR\_GAP$=120^\circ$ in our adjusted cone-based statistic. For $d=3$ and $d=5$, the same missing-fraction threshold corresponds to occupied sectors of at most $218.94^\circ$ and $206.13^\circ$, respectively.

In both boundary settings, we consider several combinations of intrinsic dimension $d$ and ambient dimension $D$.

\begin{figure}[H]
\centering
\includegraphics[scale=0.5]{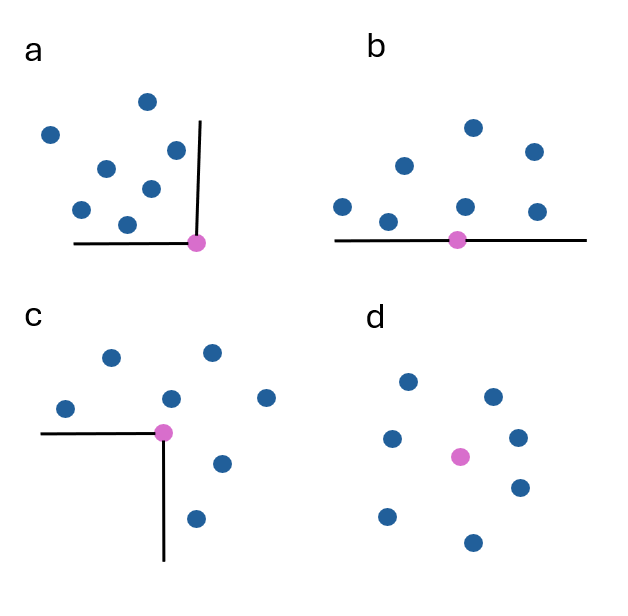}
\caption{Illustration of the local-sector dataset in $\mathbb{R}^{2}$ with $k=7$ neighbors (blue) and one focal point (pink). The occupied sectors are $90^{\circ}$ (a), $180^{\circ}$ (b), $270^{\circ}$ (c), and $360^{\circ}$ (d). For a missing fraction of at least $90^{\circ}$: $ f \geq \tau = 1/4$, the focal point in a, b and c is labeled as a boundary point, while (d) is labeled as an interior point.}
\label{fig:local-sector}
\end{figure}

\paragraph{Metrics.} Based on the two types of ground-truth measures, we compute the discrimination and the absolute error of the boundary score estimates as follows,

\begin{flalign*}
\mathrm{AUROC}&=\frac{1}
{n_1n_0}\sum_{i:y_i=1}^{n_1}\sum_{j:y_j=0}^{n_0}\left[\mathbb{I}(S_i>S_j)+\frac{1}{2}\mathbb{I}(S_i=S_j)\right],\\
\text{MAE}&=\frac{1}{n}\sum_{i=1}^{n}|y_i - \hat{y}_i|.
\end{flalign*}

The boundary score assigned by each method is denoted as $S$. As our cone-based statistics consider boundary points as points with extreme scores in both directions, 0 and 1, we define $S(x) = 2|BoS(x)-0.5|$. 

\paragraph{Hyperparameters.} In the following evaluations, the hyperplane method is supplied with the intrinsic dimension. We set PROP\_THRESH$=0.51$ and ANGULAR\_GAP$=120^{\circ}$ in the adjusted variant to increase the exploration of candidate  concave boundary points, and  DIR\_THRESH$=0.7$ in the oriented variant, corresponding to an angle of at most $\pm 45^{\circ}$ with respect to the barycenter direction.

\subsection{Results}
\paragraph{Support Boundary.}\label{sec:support-boundary-frame}

In this setting, we consider $15$ dataset families and several intrinsic $d$ and ambient $D$ dimension combinations. Mathematical definitions of these datasets and combinations of ($d,D$) are presented in Annex \ref{annex:support-dataset}. For each dataset, we fix the number of observations to $n=600$, regardless of the dimensions, and generate $5$ samples of each point cloud using different seeds. We consider $3$ neighborhood sizes: $k\in\{12,32,52\}$, we present the results for $k=32$ in this section, the two other neighborhood sizes are  displayed in Annex \ref{annex:boundary_benchmark}. Using the percentile-based rule provides for each dataset, each seed and each $k$, $n_B = 120$, $n_{\text{deep-interior}}=300$ and $n_{\text{excluded}}=180$.

The following figure shows the AUROC results averaged over the $5$ point-cloud samples. Focusing on the curse of dimensionality that occurs when increasing the dimension but keeping the number $n$ fixed, both tangent plane and hyperplane methods seem to be systematically impacted for $d=10$ whatever the number of neighbors, compared to the other statistics. The adjusted cone-based method displays as well a discrimination drop for $d=10$ but it improves across neighborhood sizes (Figure \ref{fig:support_boundary_cross} Annex \ref{annex:boundary_benchmark}). We explain that drop by the fact that, in higher intrinsic dimensions, the angular gap constraint is easily satisfied due to the density decline, hence, leading to higher boundary scores for deep-interior points compared to initial cone-based scores. The oriented cone-based statistic fails to assign a boundary score to some data points, especially for a small $k$, but neither the intrinsic nor the ambient dimension seems to impact this missing-score pattern (Figure \ref{fig:oriented-cone-based} Annex \ref{annex:support-dataset}). The AUROC computation excludes these missing scores; we do not penalize these values, which constitutes a limit in the presentation of our results, mainly for $k=12$. Among all $6$ boundary point detection methods, the oriented cone-based one is the only one to fail to assign some scores.
Overall, we observe the competitive behavior of our cone-based statistics with an increase of the discrimination performance for higher $k$-nn.

\begin{figure}[H]
\centering
\includegraphics[scale=0.175]{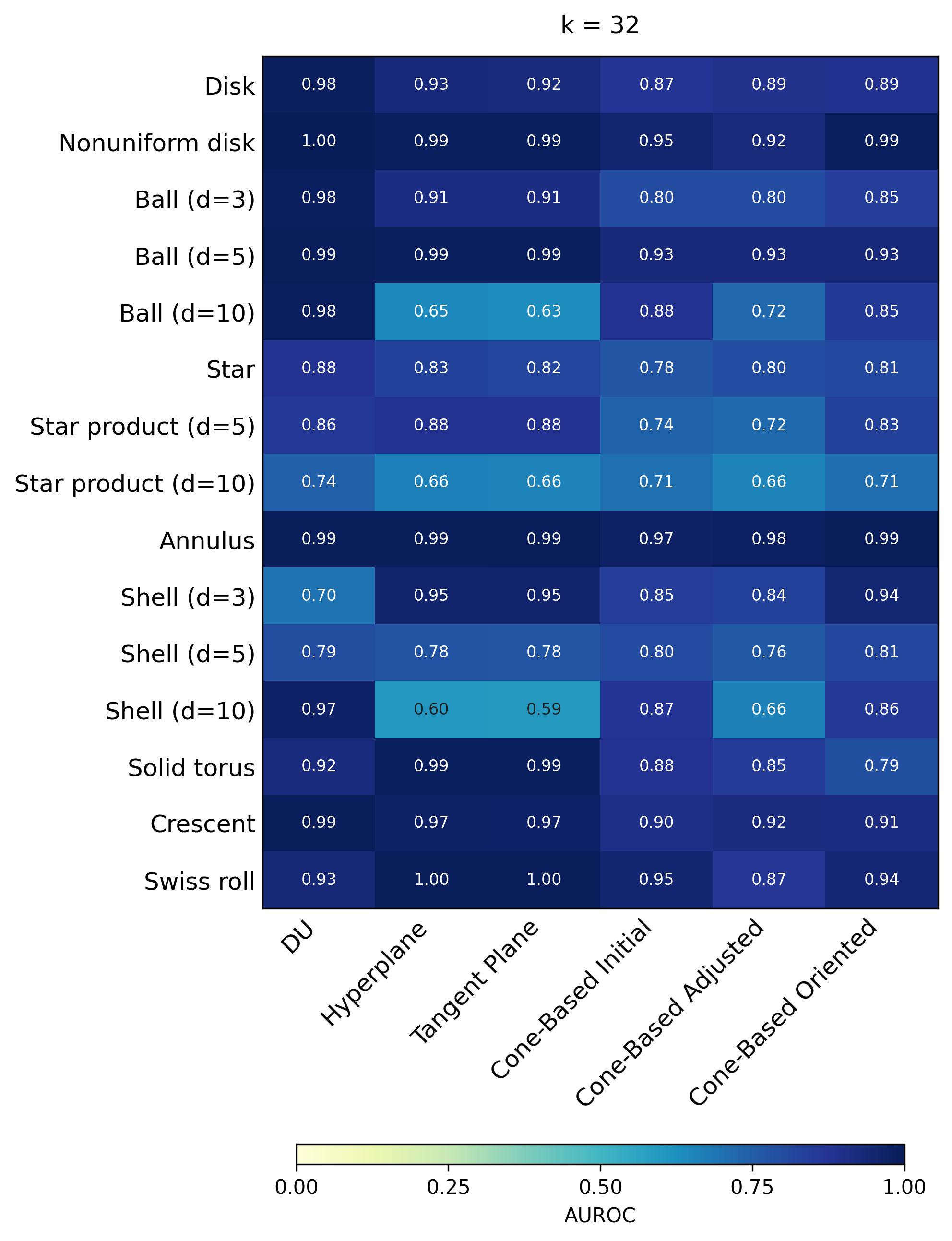}
\caption{AUROC results for the $3$ SOTA methods and our cone-based statistics, for $k=32$. As expected from isometric embeddings, no score variability was observed when increasing $D$ and keeping the sample fixed, hence we only present AUROC results with respect to $d$.}
\label{fig:support-bde}
\end{figure}

\paragraph{Local-Sector Boundary.}

This second setting defines a boundary point via the local configuration of its neighborhood. 
The benchmark considers $7$ full angular openings $\alpha$ of the occupied spherical caps,
\[
\alpha \in \{60^\circ,90^\circ,120^\circ,180^\circ,
240^\circ,300^\circ,360^\circ\}.
\]
For the missing-fraction threshold $\tau=1/3$, the first $5$ configurations are labeled as boundary points for $d=2$, whereas only the first $4$ are labeled as boundary points for $d\in\{3,5\}$. The remaining configurations are labeled as interior points. For each $\alpha$, $10$ focal-point scenarios are independently generated, providing a total number of  $70$ focal points per local sector ($d$,$D$).

\begin{figure}[H]
\centering
\includegraphics[scale=0.175]{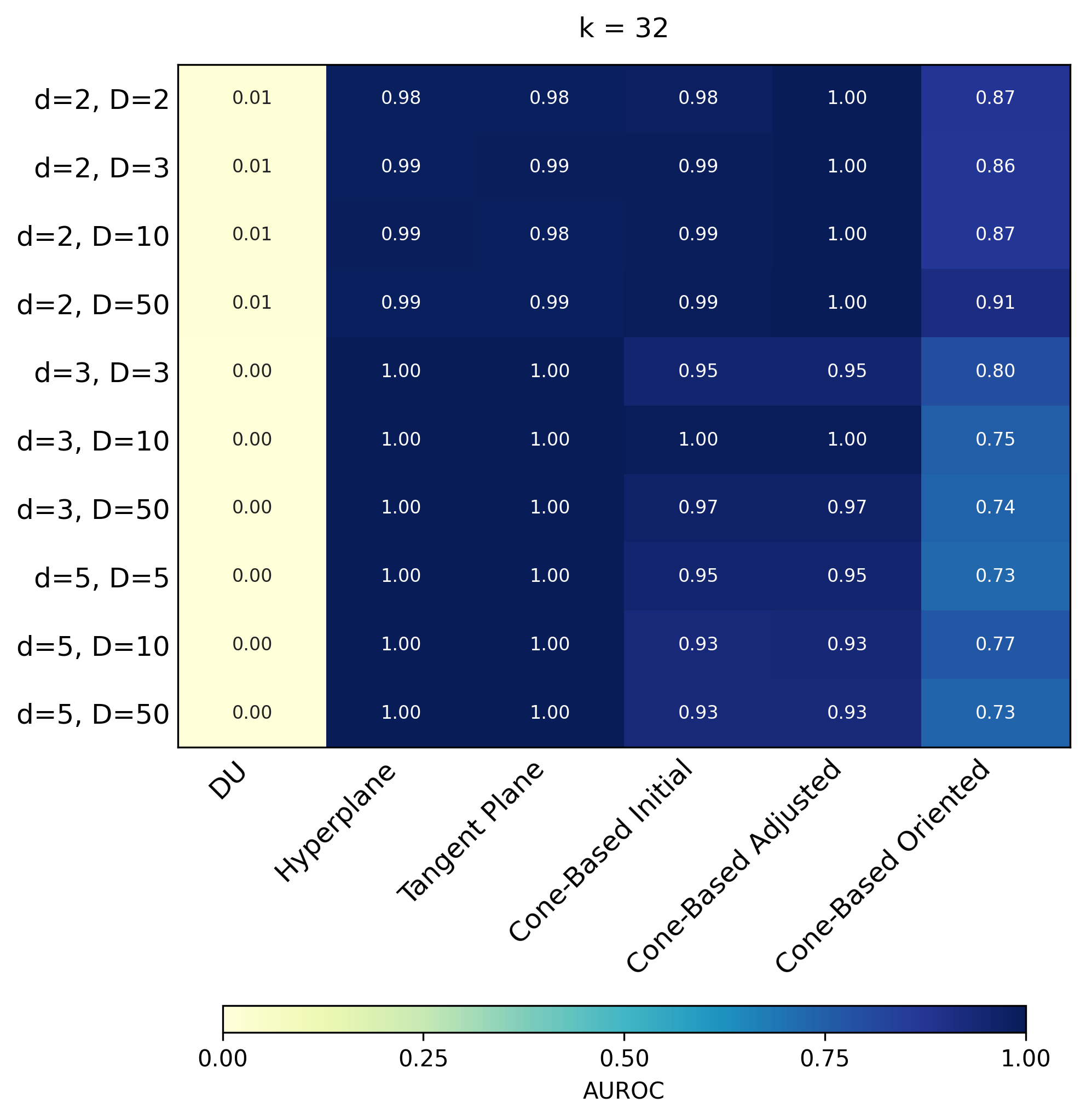}
\caption{AUROC results of the focal points with $k=32$. See Figure \ref{fig:support_boundary_cross} Annex \ref{annex:boundary_benchmark} for the results in $k=\{12,52\}$.}
\label{fig:heatmap-local-sec}
\end{figure}

From Figure \ref{fig:heatmap-local-sec}, we observe an AUROC near $0$ for the DU estimate. This is explained by the inverted ranking provided by that method; AUROC(-DU) = 1-AUROC(DU). In this focal point setting, $x_i$ points in the opposite direction compared to its neighbors, therefore, the sum of all dot products, $\sum_{j \in K} <v_j, v_i>$ is negative when $x_i$ is a boundary point (\cite{hickok2026boundarydetectionhighdimensionalpoint}). Inverting the sign through AUROC(-DU) = 1-AUROC(DU) offers a high discrimination (Figure \ref{fig:inverted_DU} Annex \ref{annex:boundary_benchmark}).

The other statistics provide high discrimination performance as well, except for the oriented cone-based one which is explainable by its barycenter alignment. Given the missing fraction $\tau = 1/3$ that we use, the oriented cone points towards the barycenter which implies a convex cone and the projection of some neighbors outside of the latter cone. We report the proportion of focal points whose oriented-boundary scores are missing (Figure \ref{fig:oriented_nan_local} Annex \ref{annex:support-dataset}). In this evaluation setting, the increase of $k$ does not lower the number of missing scores.
Lastly, even though the adjusted cone-based statistic is computed in the span of the two vector basis, it seems to discriminate well in high intrinsic dimensions $d=\{3,5\}$, especially for $k=32$.

To conclude this analysis, we underline that the generalizability of these findings is limited to the synthetic local-sector configurations considered here, where neighbor directions are sampled uniformly within a single spherical cap. More complex directional distributions such as non-uniform angular densities within the occupied region and distributions perturbed by noise or outliers, are not evaluated here.

\paragraph{Adjusted Cone-based Statistic for Concavity.}

We test here the concavity information capture of our adjusted cone-based estimate in $\mathbb{R}^{2}$, compared to both other cone-based methods. Following the local-sector dataset construction, we consider $11$ non-zero angular gaps among $[30^{\circ}, 330^{\circ}]$ in steps of $30^{\circ}$. Angular gaps (empty sector openings) from $30^{\circ}$ until $150^{\circ}$ correspond to \textit{concave} cases, gaps from $210^{\circ}$ until $330^{\circ}$ relate to \textit{convex} cases, and gaps of $180^{\circ}$ and $0^{\circ}$ define \textit{flat} and \textit{interior} cases, respectively. Each empty-sector configuration among the $12$ ones is generated 100 times with different seeds and the boundary score is computed only for the focal points. We estimate the angular gaps using all three cone-based statistics, expecting that the adjusted variant will outperform the other two.

 Concavity, convexity and flat cases are defined for \textit{final} BoS\_THRESH$=0.7$, i.e., $BoS(x)\geq0.7 \cup BoS(x)\leq 0.3$. Final refers to the scores and angles after adjustment or fall back to the initial values if no adjustment, the oriented $BoS(x)$ is already the final score in its category. Points whose final $BoS(x) \in (0.3,0.7)$ are considered as interior points and their angular gaps are then set to $0^{\circ}$.  We review initial cone aperture for adjustment when $BoS(x)\geq0.51 \cup BoS(x)\leq 0.49$, corresponding to PROP\_THRESH$=0.51$. The cone opening is adjusted when ANGULAR\_GAP $\geq 30^{\circ}$. For points associated with final $BoS(x)\geq 0.7$, the estimated empty sector is given by $360^{\circ}-\theta$. For points having final $BoS(x)\leq 0.3$, the empty sector is given by $\theta$.

 We display the classification results only for the concavity and convexity cases in the following figure as our method does not allow angles of $180^{\circ}$ to avoid co-linearity of the basis vectors; the estimated angles of the occupied sector will be close to $180^{\circ}$ but slightly lower. However, we present the MAE values in all four cases to collect the error magnitude.

\begin{figure}[H]
\centering
\includegraphics[scale=0.1]{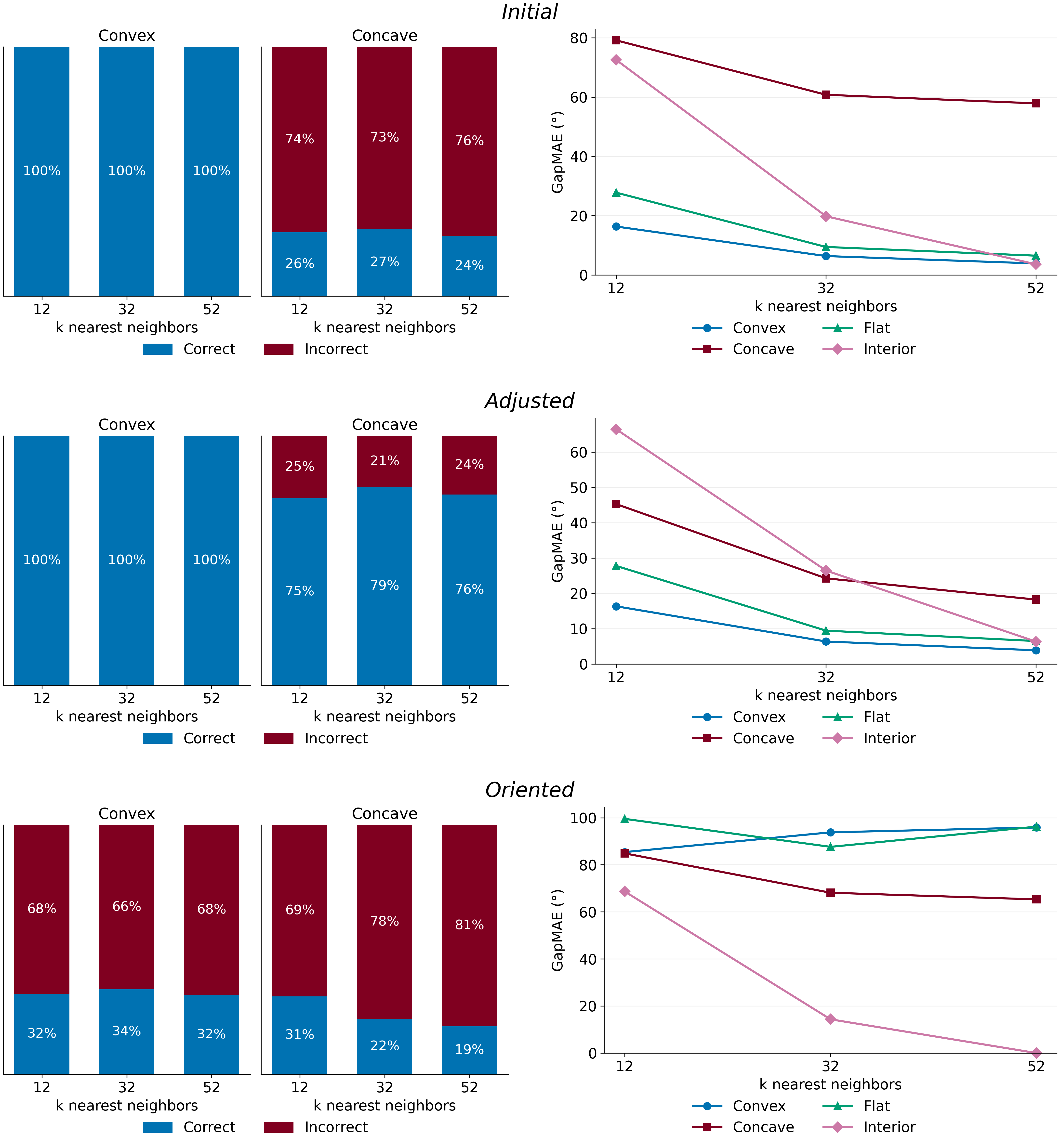}
\caption{classification of convexity and concavity cases (left) and MAE between the true angular gap and its cone-based estimate (right).}
\label{fig:oncavity-cross-051}
\end{figure}

 From Figure \ref{fig:oncavity-cross-051}, we observe that, as expected, adjusting the initial cone aperture provides better concavity classifications and smaller errors (y-axis graduation is at most $50^{\circ}$) in the estimation of the true angular gap. The oriented statistic provides the worst performance in assigning the correct classification and gap degree, explainable by the constraint on the alignment towards the barycenter.

This experiment shows the benefit of adjusting the initial cone opening compared to the initial and oriented variants. In the future, we envisage the comparison with direct maximum angular-gap estimators and the validation
of concavity characterization in higher intrinsic dimensions.

\section{RAG-related Application}\label{sec:application}

We propose to use the local geometric information provided by our methods to reduce embedding dimensions. Embedding reduction techniques are widely used in Retrieval-Augmented Generation (RAG) research (\cite{salama-etal-2024-semantic}; \cite{khaledian2025pcaragprincipalcomponentanalysis}; \cite{math13152469}; \cite{11605980}; \cite{sun2026svd}). RAG workflows consist in retrieving texts similar to a query in large document collections. All texts are embedded in high-dimensional vectors, \textit{embeddings}, which implies the need of storage resources and a certain latency when searching among this high-dimensional point cloud (\cite{huergapérez2025optimizationembeddingsstoragerag}). Therefore, we suggest to apply our cone-based statistics to reduce embeddings' dimensions and assess their effects on positive–negative cosine-similarity margins relative to the original embedding space. As PCA is still widely used in that field (\cite{khaledian2025pcaragprincipalcomponentanalysis}; \cite{math13152469}; \cite{huergapérez2025optimizationembeddingsstoragerag}) and can benefit from pre-processing strategies  (\cite{raunak-etal-2019-effective}), we consider two PCA-based approaches to find lower-dimensional representations: retrieving principal directions from boundary points and scaling the observations with their boundary scores before computing the covariance matrix. We motivate the first approach by the intuition that boundary points contain most of the point cloud variance, hence would provide principal directions similar to the standard PCA while minimizing the training sample  (\cite{ALAM2020268}). The second strategy uses all the boundary-score information to scale vectors' importance, hypothesizing that boundary points describe more significant patterns than interior points.

\paragraph{Dataset.} We choose the fact-checking dataset SciFact from \href{https://huggingface.co/datasets/BeIR/beir-corpus}{HuggingFace BEIR datasets} (\cite{NEURIPSBEIR}) that associates texts from \textcite{wadden-etal-2020-fact} with corresponding queries. Hence, for each query, the relevant text(s) is known. The query set (train and test sets) is composed of $1,109$ observations and the text set of $5,183$. All texts and queries are embedded using the BGE-M3 model (\cite{bge-m3}) that produces $1024$-dimensional embeddings with an input window up to $8,192$ tokens. This high token window avoids us to chunk text as the SciFact longest one is up to $1,923$ tokens. In that $1024$-dimensional  space, we compute the cosine similarity between each query and its matching text(s). In this dataset, one query $i$ can have up to $5$ matching texts $j$, we denote it as $j \in \mathcal{T}_i$. The total amount of pairs is $1,258$. These annotated relevant query–document pairs constitute the positive sample set and we compare it with $5$ randomly selected negative samples of query and non-matching text pairs. This pointwise positive-negative sample comparison is denoted as

\begin{equation*}
    \Delta^{+-_{l}}_{i,j} = \cos(q_i,t_{i,j}^{+}) - \cos(q_i,t_{i,j}^{-}),
\end{equation*}
with $i =\{1,...,1109\}$, $j\in \mathcal{T}_i$, and $l=\{1,...,5\}$ referring to the index of the negative sample. The ideal scenario is to obtain $\Delta^{+-_{l}}_{i,j}>0\ \forall i,j,l$.
This delta, or \textit{margin}, is computed in the original embedding space and in each PCA-based reduced embedding space.

\paragraph{Boundary-guided PCA.} We consider $7$ PCA-based approaches and $5$ PC dimensions $D_{\text{Reduced}}=\{16,32,64,128,256\}$. The first approach is the standard PCA applied using all text embeddings. The second, third and fourth ones, \textit{$v$-Boundary PCA}, consist of applying the PCA uniquely on boundary points computed with the variant $v=\{\text{initial, adjusted, oriented}\}$. The last three approaches use all persistent-boundary scores (eq. \ref{eq:persistent-bound}) to scale the observations before computing the covariance matrix and apply the PCA. We denote them as \textit{$v$-scaled PCA}. For both PCA-based approaches, we suggest to sample boundary scores over the $14$ projection dimensions $d_{B}=\{2,...,15\}$ and compute the persistent-boundary score defined as, 

\begin{equation*}\label{eq:persistent-bound}
    P^{v}_i = \frac{1}{14} \sum_{d_{B}=2}^{15} 2|BoS^{(v,d_{B})}(x_i)-0.5|.
\end{equation*}

This persistent score measures the tendency of a point to be a boundary ($P_i^{v}$ near $1$) or an interior point ($P_i^{v}$ near $0$). For instance, setting a threshold of $P^{v}_i \geq 0.8$ encompasses points with, on average, $BoS^{(v,d_{B})}(x_i) \leq 0.1$ and/or $BoS^{(v,d_{B})}(x_i) \geq 0.9$.
Each $BoS^{(v,d_{B})}(x_i)$ is computed using $32$ nearest neighbors, the ANGULAR\_GAP is set to $90^{\circ}$ and PROP\_THRESH to $0.51$ in the adjusted method. The oriented cone-based score uses the alignment constraint of $0.7$, i.e., an angle gap of at most  $\pm 45^{\circ}$ with the direction towards the barycenter. Figure \ref{fig:persistent_bos_variant} Annex \ref{annex:application} displays the persistent-boundary score distribution per variant. In the context of $v$-boundary PCAs, we set a persistent threshold for each variant such the that the number of boundary points in each variant set is greater than $\max(D_{\text{Reduced}})$ (see Annex \ref{annex:application} for motivational details). Regarding the $v$-scaled PCAs, we use $P_i^{v}$ to scale the $1024$-dimensional embeddings $x_i$. The mean and covariance matrix of the scaled observations are given by,
\begin{equation*}
\mu_P = \frac{1}{N}\sum_{i=1}^{N} P_i x_i, \quad 
\Sigma_P = \frac{1}{N-1}\sum_{i=1}^{N}
\left(P_i x_i-\mu_P\right)
\left(P_i x_i-\mu_P\right)^{\top}.
\end{equation*}

This scaling approach differs from standard observation-weighted PCA. While the latter uses weights to determine each observation’s contribution to the mean and covariance estimates, the one we apply here preserves the position of boundary points ($P_i^{v}$ near $1$) but sends interior points closer to the origin.

 For all $6$ boundary-guided PCAs, only the text embeddings are used for the centering and the collection of PCA axis. The $v$-boundary PCAs use a subset of the $5$k text embeddings, corresponding to persistent-boundary text embeddings. The $v$-scaled PCAs scale all text embeddings. The queries are then centered by the mean specific to each PCA and projected on the retrieved axis to finally compute the cosine similarities in the reduced space.

\paragraph{Metrics.} In addition to the measure $\Delta^{+-_{l}}_{i,j}$, we are interested in the impact of the dimensions' reduction on the cosine similarities, hence we compute two statistics that compare the cosine similarity distribution $\Delta^{+-}$ in the full $1024$-dimensional space with the one in each PCA-based reduced space $R$ of dimensions $\in D_{\text{Reduced}}$ that we denote $R(d)$, with $R\in \{\text{raw PCA}, \text{$v$-boundary PCA}, \text{$v$-scaled PCA}\}$. As $|D_{\text{Reduced}}|=5$ and $|R|=7$, we consider $35$ reduced spaces to compare with the original one. 
The first comparison measure is the paired margin mean expressed in cosine similarity units and defined as 

\begin{equation*}
    \widehat{\text{Margin}^{+-}}(R(d)) = \frac{1}{1,258\times 5} \sum_{i=1}^{1,109}\sum_{j\in \mathcal{T}_i}\sum_{l=1}^{5} \Delta^{+-_{l}}_{i,j}(R(d)) - \Delta^{+-_{l}}_{i,j}(1024). 
\end{equation*}

The second metric we compute is the win-rate change average expressed in percentage-point (pp) and defined as 

\begin{equation*}
    \widehat{\text{Win}^{+-}}(R(d)) = \frac{1}{5} \sum_{l=1}^{5}\frac{100}{1,258} \left(\#(\Delta^{+-_{l}}_{i,j}(R(d))>0) - \#(\Delta^{+-_{l}}_{i,j}(1024)>0)\right).
\end{equation*}

Both statistics aggregate changes across query–document comparisons, hence, when comparing reduced PCA-based statistics with the original embedding space, we capture overall alignment and not individual similarity margin comparisons.
The best-case scenario would be to preserve or enhance the clear separation between both positive and negative cosine similarities when reducing the dimensions which translates to 
$\widehat{\text{Margin}^{+-}}\geq0$ and $\widehat{\text{Win}^{+-}}\geq0$. 
Finally, we emphasize that the comparison between each PCA-based approach and the original (uncentered) cosine similarities encompasses two impacts: the dimension reduction but also the centering effect; each PCA-based technique performs a centering of the embeddings which is unique to it (dependent on the selected/scaled data). We decide to use the non-centered original cosine similarities as the common reference to assess each PCA-based overall performance; in a RAG workflow it seems more natural to pass from the original space to the reduced one without performing a centering specific to the reduction method.

\begin{figure}[H]
    \centering
    \includegraphics[width=1\linewidth]{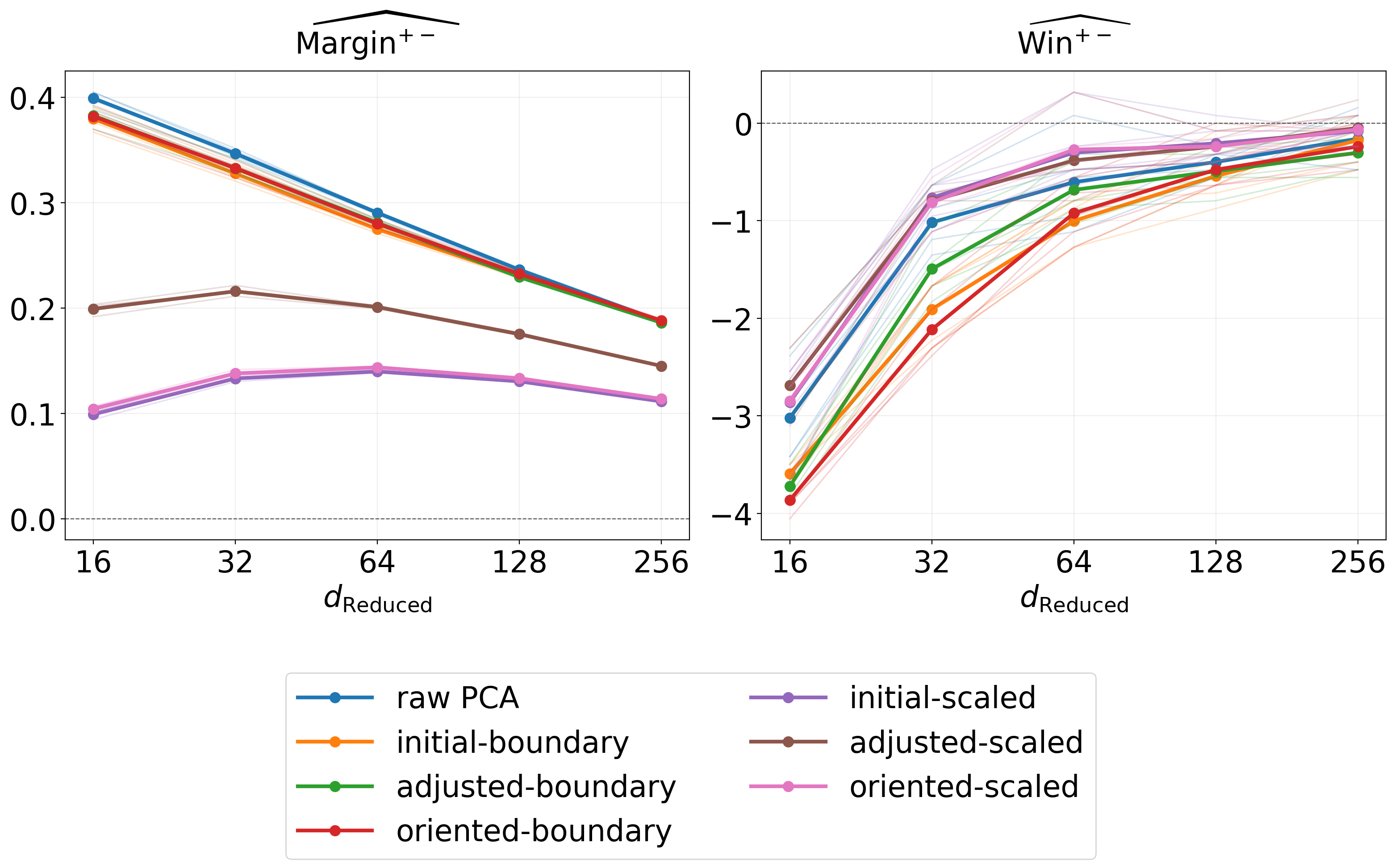}
    \caption{Bold curves represent the mean over $5$ negative samples and thin curves represent each of these samples. Left: the y-axis expresses difference in positive-negative cosine margins between the original embedding space and each 7 reduced counterparts. Right: the y-axis represents the pp difference with respect to the original space; $0$ pp means that the win-rate is preserved when reducing the dimensions and $-4$ pp means that the win-rate decreases by $4$ pp during the dimension reduction.}
    \label{fig:application_results}
\end{figure}

\paragraph{Results.} Focusing on $\widehat{\text{Margin}^{+-}}$ measures (Figure \ref{fig:application_results}, left), we see that the raw (standard) PCA and the 3 $v$-boundary PCAs provide a greater polarization between positive and negative cosine similarities, especially for small $D_{\text{Reduced}}$, compared to the cosine similarities in the original space. On the contrary, the $3$ $v$-scaled PCAs provide a similar polarization as in the original space. Regarding the win-rate measure (right), we observe more similar win-rates between the $v$-scaled PCAs and the original one, compared to the $v$-boundary PCAs. The raw PCA shows more of an intermediate pattern. There seems to be a trade-off between the polarization and the win-rate: the PCA-based approaches providing better polarizations, give smaller win-rate compared to the original space.

Accounting for this trade-off, we observe that scaling the embeddings with their boundary-persistent scores shows better retention of both selected aggregate measures than boundary-subset PCA in this experiment. Comparing the variant effects, it appears that both initial and oriented boundary scores provide more aligned polarization and win-rate with respect to the original space. As these results encompass both the centering and the dimension reduction effects of each approach, it would be interesting to isolate the effect of each factor for each boundary-guided PCA in future work.

We underline the variability of win-rate curves among the $5$ negative samples, which constitutes a limit to the generalization of our results with respect to this measure. Finally, we restricted the comparison framework of reduction techniques to the PCA, but other methods such as Kernel PCA (\cite{11605980}) or Discrete Wavelet Transform (\cite{salama2025combining}; \cite{salama-etal-2024-semantic}) might offer other insights with respect to our PCA-based approaches.

\section{Conclusion}

Benchmarking our three cone-based variants against three SOTA methods (\cite{tangent}; \cite{hickok2026boundarydetectionhighdimensionalpoint}) in noiseless synthetic datasets, reveals competitive performance in both boundary settings, including at higher intrinsic dimensions, where the range of possible cone orientations increases.
Compared to the selected SOTA methods applicable in high-dimensions, our cone-based statistics additionally assign a local angular configuration to candidate boundary points, distinguishing concave and convex configurations.

Adjusting the initial cone aperture to account for the concavity information provides better estimations of the angular gap, or empty sector opening, and better detections of concavity in $2$ dimensions. The latter method also generalizes well as a boundary point detector in high-dimensional local sectors. The oriented cone-based method sometimes fails to assign boundary scores, especially when the number of neighbors is small ($k=12$), but this issue tends to disappear with $k=\{32,52\}$ in the support-boundary setting. The evaluation of this oriented statistic as a boundary-relative centrality measure will be investigated in future work, along with the robustness of our methods in noisy spaces and their computational costs, especially for the adjusted variant. We will also report on their statistical properties such as (asymptotic) consistency and convergence rate.

The application of our cone-based statistics to embedding dimensionality reduction reveals different degrees of alignment with the original embedding space according to the two margin-based statistics. Although we restrict the experiment to one dataset and one embedding model, our results suggest that scaling embeddings via their persistent-boundary scores before estimating the covariance matrix, rather than restricting PCA to candidate boundary points, better preserves the evaluated cosine-similarity-based statistics of the original embedding space.

Python scripts used in the benchmark evaluations are available on \href{https://github.com/lynndelcon/A-Cone-Based-Local-Geometric-Statistic-for-Identifying-Candidate-Boundary-Points-and-their-Concavity}{GitHub}.

\section*{Acknowledgments}
This research was supported by the Onderzoeksprogramma Artifici\"ele Intelligentie (AI) Vlaanderen program.
DT acknowledges support from the Research Foundation–Flanders (FWO) via the Odysseus II programme no. G0DBZ23N. AA acknowledges support from the Vrije Universiteit Brussel Research Council (VUB-OZR), the Francqui Foundation (Belgium) through a Francqui Start-Up Grant, and a fellowship from the Research Foundation Flanders (FWO) under Grant No.1286924N. VG acknowledges support from Research Foundation Flanders under Grant No.G032822N and G0K9322N.  
The computational resources and services used in this work were provided by the VSC (Flemish Supercomputer Center), funded by the Research Foundation Flanders (FWO) and the Flemish Government - department WEWIS.

\section*{AI-Use Disclosure}

The proposed boundary-detection algorithm and its variants were conceptualized and developed in pseudocode without AI assistance. GPT-5.5 (Medium Thinking) was used for code syntax verification and minor corrections. GPT-6.1-Sol (High Thinking) was used to implement the synthetic benchmark datasets based on mathematical definitions provided by the authors and to provide feedback on the manuscript. GPT-6-Astra (Low Thinking) implemented in python the application analysis.
All AI-assisted implementations and analyses were reviewed and validated by the authors, who retain full responsibility for the scientific content and conclusions of this work.

\newpage
\printbibliography
\newpage
\appendix

\section{Support-Boundary Dataset}\label{annex:support-dataset}

For a point $x$ in a known support $\Omega\subseteq\mathbb{R}^d$, the
reference distance to its boundary is
\begin{equation*}
  d_{\partial\Omega}(x)
  :=\operatorname{dist}(x,\partial\Omega)
  =\inf_{y\in\partial\Omega}\|x-y\|_2.
  \label{eq:support-distance-general}
\end{equation*}
Except for the Swiss roll (whose benchmark target is an \emph{intrinsic}
manifold-boundary distance), all distances below are Euclidean distances in
the intrinsic coordinate space.\\

As we consider several combinations of intrinsic ($d$) and ambient ($D$) dimensions, we increase the ambient dimension as follows. If a dataset in $\mathbb R^d$ is mapped to $\mathbb R^D$ ($D\geq d$)
by $y=Qx$, where $Q\in\mathbb R^{D\times d}$ satisfies
$Q^\top Q=I_d$, then in the absence of noise,
\begin{equation*}
  \operatorname{dist}_{\mathbb R^D}
  \bigl(Qx,Q(\partial\Omega)\bigr)
  =\operatorname{dist}_{\mathbb R^d}(x,\partial\Omega).
\end{equation*}

Consequently, Euclidean distances, angles, and nearest-neighbor rankings are preserved under this noiseless isometric embedding. The benchmark therefore assesses the sensitivity of boundary detection methods to increasing ambient dimensions without introducing additional geometric information or noise.

\begin{table}[H]
\centering
\caption{Support-boundary benchmark datasets and their intrinsic ($d$) and ambient ($D$) dimensions. Each configuration contains $n=600$ points. }
\label{tab:support_dataset_dimensions}
\begin{tabular}{lcl}
\hline
Dataset family & Intrinsic dimension $d$ & Ambient dimensions $D$ \\
\hline
Disk & 2 & 2, 3, 10, 50 \\
Nonuniform disk & 2 & 2, 3, 10, 50 \\
Ball ($d=3$) & 3 & 3, 10, 50 \\
Ball ($d=5$) & 5 & 5, 10, 50 \\
Ball ($d=10$) & 10 & 10, 50 \\
Star & 2 & 2, 3, 10, 50 \\
Star product ($d=5$) & 5 & 5, 10, 50 \\
Star product ($d=10$) & 10 & 10, 50 \\
Annulus & 2 & 2, 3, 10, 50 \\
Shell ($d=3$) & 3 & 3, 10, 50 \\
Shell ($d=5$) & 5 & 5, 10, 50 \\
Shell ($d=10$) & 10 & 10, 50 \\
Solid torus & 3 & 3, 10, 50 \\
Crescent & 2 & 2, 3, 10, 50 \\
Swiss roll & 2 & 3, 10, 50 \\
\hline
\end{tabular}
\end{table}

The star product refers to the product of the five-lobed radial star with the interval $[-1,1]^{d-2}$, and the shell is the higher-dimensional generalization of the annulus.\\

\textbf{Unit disk.}
\begin{flalign*}
  &\Omega_{\mathrm{disk}}
    =\{x\in\mathbb{R}^2:\|x\|_2\leq 1\},\\
  &d_{\partial\Omega_{\mathrm{disk}}}(x)
    =1-\|x\|_2.
\end{flalign*}
The benchmark samples $r=\sqrt U$ and $\theta\sim\mathcal{U}(0,2\pi)$,
where $U\sim\mathcal{U}(0,1)$ and $x=r(\cos\theta,\sin\theta)$.\\

\textbf{Non-uniform unit disk.}
The support and boundary distance are identical to those of the unit disk,
\begin{flalign*}
  &\Omega_{\mathrm{disk,nu}}=\Omega_{\mathrm{disk}},\\
  &d_{\partial\Omega_{\mathrm{disk,nu}}}(x)=1-\|x\|_2.
\end{flalign*}
Only the sampling law changes: with probability $0.75$, points are sampled
uniformly in the disk of radius $0.55$; with probability $0.25$, they are
sampled uniformly in the annulus with radii $0.55$ and $1$.  This tests
sensitivity to density variation without changing the support boundary.\\

\textbf{Annulus.}
\begin{flalign*}
  &\Omega_{\mathrm{ann}}
    =\{x\in\mathbb{R}^2:1\leq\|x\|_2\leq2\},\\
  &d_{\partial\Omega_{\mathrm{ann}}}(x)
    =\min\{\|x\|_2-1,\;2-\|x\|_2\}.
\end{flalign*}
The two terms refer to the inner and outer circular boundaries, respectively.\\

\textbf{Five-lobed radial star.}
Write $u(\theta)=(\cos\theta,\sin\theta)$ and
$R(\theta)=1+a\cos(m\theta)$, with $a=0.32$ and $m=5$.
\begin{flalign*}
  &\Omega_{\mathrm{star}}
    =\{r\,u(\theta):0\leq r\leq R(\theta),\;
      \theta\in[0,2\pi)\},\\
  &\partial\Omega_{\mathrm{star}}
    =\{R(\theta)u(\theta):\theta\in[0,2\pi)\},\\
  &d_{\partial\Omega_{\mathrm{star}}}(x)
    =\min_{\theta\in[0,2\pi)}
      \|x-R(\theta)u(\theta)\|_2.
\end{flalign*}
The benchmark approximates this minimization by nearest-neighbor search
against $30{,}000$ uniformly spaced samples of the analytical boundary
curve; thus its stored distance is numerical, not an exact minimizer.\\

\textbf{Crescent.}
Let $R=1.5$, $r=1.15$, and $c=(0.60,0)$.  The support is a disk
with an overlapping offset disk removed,
\begin{equation*}
  \Omega_{\mathrm{cres}}
    =\{x\in\mathbb{R}^2:\|x\|_2\leq R,
      \ \|x-c\|_2\geq r\}.
\end{equation*}
Its boundary consists of the \emph{valid arcs}
\begin{flalign*}
  &\Gamma_{\mathrm{out}}
    =\{y:\|y\|_2=R,\ \|y-c\|_2\geq r\},\\
  &\Gamma_{\mathrm{in}}
    =\{y:\|y-c\|_2=r,\ \|y\|_2\leq R\},\\
  &d_{\partial\Omega_{\mathrm{cres}}}(x)
    =\min\{\operatorname{dist}(x,\Gamma_{\mathrm{out}}),
              \operatorname{dist}(x,\Gamma_{\mathrm{in}})\}.
\end{flalign*}
The benchmark approximates these distances using densely sampled valid
arcs (up to $30{,}000$ angular samples per circle).\\

\textbf{Solid torus.}
For $x=(x_1,x_2,x_3)\in\mathbb{R}^3$, define
\begin{equation*}
  q(x)=\sqrt{\bigl(\sqrt{x_1^2+x_2^2}-R\bigr)^2+x_3^2},
  \qquad R=2,\quad r=0.7.
\end{equation*}
Then
\begin{flalign*}
  &\Omega_{\mathrm{tor}}=\{x\in\mathbb{R}^3:q(x)\leq r\},\\
  &d_{\partial\Omega_{\mathrm{tor}}}(x)=r-q(x).
\end{flalign*}
The distance is to the surface of the solid tube around a circle of
major radius $R$ and minor radius $r$.\\

\textbf{Swiss-roll manifold.}
This is a two-dimensional manifold embedded in $\mathbb{R}^3$, parametrized by
\begin{flalign*}
  &F(t,z)=(t\cos t,\ z,\ t\sin t),\\
  &\Omega_{\mathrm{roll}}
    =F\bigl([t_0,t_1]\times[-2,2]\bigr),\\
  &t_0=1.5\pi,\qquad t_1=4.5\pi.
\end{flalign*}
The induced metric is
$\mathrm{d}s^2=(1+t^2)\,\mathrm{d}t^2+\mathrm{d}z^2$.
Define the arc-length coordinate
\begin{equation*}
  A(t)=\int_{t_0}^{t}\sqrt{1+u^2}\,\mathrm{d}u
  =\frac12\Bigl[t\sqrt{1+t^2}+\operatorname{arsinh}(t)
      -t_0\sqrt{1+t_0^2}-\operatorname{arsinh}(t_0)\Bigr].
\end{equation*}
The benchmark's target distance is the intrinsic (geodesic) distance
from $F(t,z)$ to the manifold boundary
\begin{equation*}
  d_{\partial\Omega_{\mathrm{roll}}}^{\mathrm{intr}}(F(t,z))
   =\min\bigl\{A(t),\ A(t_1)-A(t),\ z+2,\ 2-z\bigr\}.
\end{equation*}
The code approximates $A(t)$ by numerical integration on a grid of
$30{,}000$ points.  This distance is \emph{not} the shortest ambient
Euclidean distance from $F(t,z)$ to its boundary curves.\\

\textbf{High-dimensional unit balls.}
For $d\in\{3,5,10\}$,
\begin{flalign*}
  &\Omega_{\mathrm{ball}}^{(d)}
    =\{x\in\mathbb{R}^d:\|x\|_2\leq1\},\\
  &d_{\partial\Omega_{\mathrm{ball}}^{(d)}}(x)
    =1-\|x\|_2.
\end{flalign*}
Sampling uses a uniformly distributed direction on $\mathbb S^{d-1}$
and radius $U^{1/d}$, with $U\sim\mathcal U(0,1)$.\\

\textbf{High-dimensional shells.}
For $d\in\{3,5,10\}$ and $a=0.65$,
\begin{flalign*}
  &\Omega_{\mathrm{shell}}^{(d)}
    =\{x\in\mathbb{R}^d:a\leq\|x\|_2\leq1\},\\
  &d_{\partial\Omega_{\mathrm{shell}}^{(d)}}(x)
    =\min\{\|x\|_2-a,\ 1-\|x\|_2\}.
\end{flalign*}
Sampling uses a uniform direction and radial coordinate
$\bigl(a^d+(1-a^d)U\bigr)^{1/d}$, with $U\sim\mathcal U(0,1)$.\\

\textbf{High-dimensional star products.}
For $d\in\{5,10\}$, let $x=(u,z)$ with
$u\in\mathbb R^2$ and $z=(z_1,\ldots,z_{d-2})\in\mathbb R^{d-2}$.
\begin{flalign*}
  &\Omega_{\mathrm{star\text{-}prod}}^{(d)}
    =\Omega_{\mathrm{star}}\times[-1,1]^{d-2},\\
  &d_{\partial\Omega_{\mathrm{star\text{-}prod}}^{(d)}}(u,z)
    =\min\left\{
        d_{\partial\Omega_{\mathrm{star}}}(u),\;
        \min_{1\leq j\leq d-2}(1-|z_j|)
      \right\}.
\end{flalign*}
The star component uses the numerical boundary-distance approximation
described above; the hypercube-face distances are exact.\\

\textbf{Isometric ambient embeddings (all support families).}
If a dataset in $\mathbb R^d$ is mapped to $\mathbb R^D$ ($D\geq d$)
by $y=Qx$, where $Q\in\mathbb R^{D\times d}$ satisfies
$Q^\top Q=I_d$, then in the absence of noise,
\begin{equation*}
  \operatorname{dist}_{\mathbb R^D}
  \bigl(Qx,Q(\partial\Omega)\bigr)
  =\operatorname{dist}_{\mathbb R^d}(x,\partial\Omega).
\end{equation*}
For the Swiss roll, the inherited label remains the intrinsic distance
specified above.

\begin{figure}[H]
    \centering
    \includegraphics[width=0.7\linewidth]{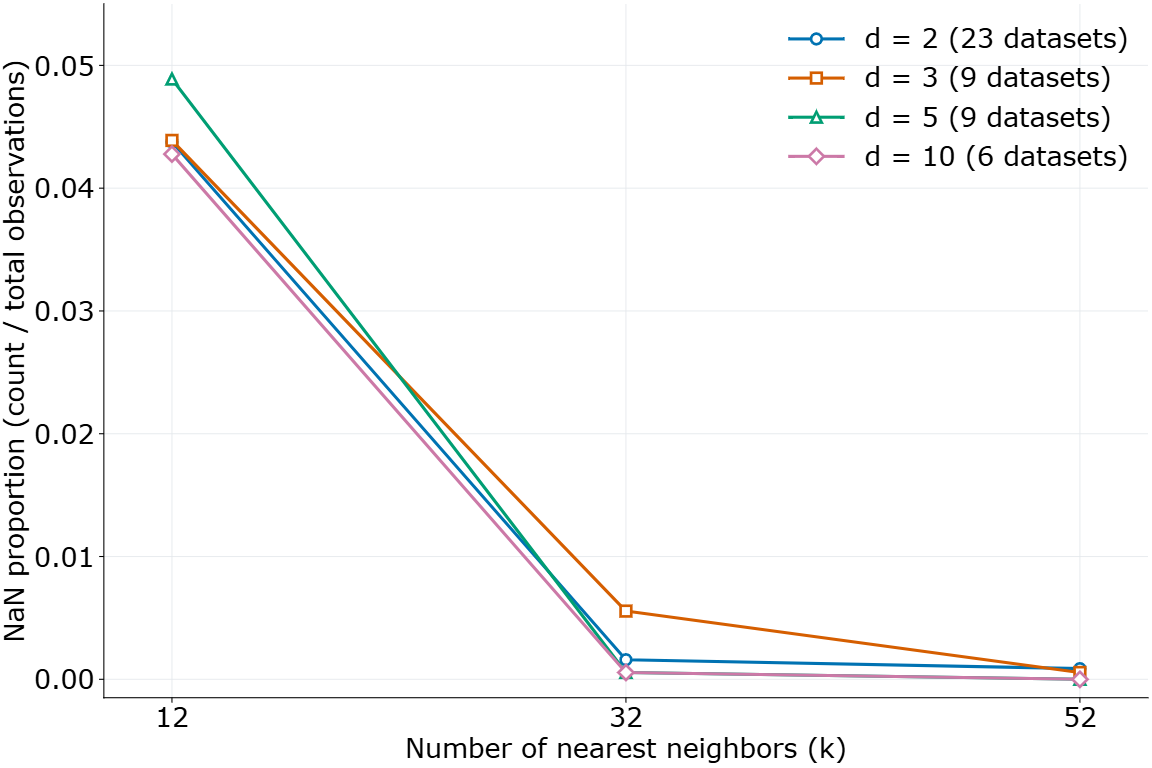}
    \caption{Proportion of missing scores from the oriented cone-based statistic in the support-boundary evaluation setting.}
    \label{fig:oriented-cone-based}
\end{figure}

\begin{figure}[H]
    \centering
    \includegraphics[width=0.7\linewidth]{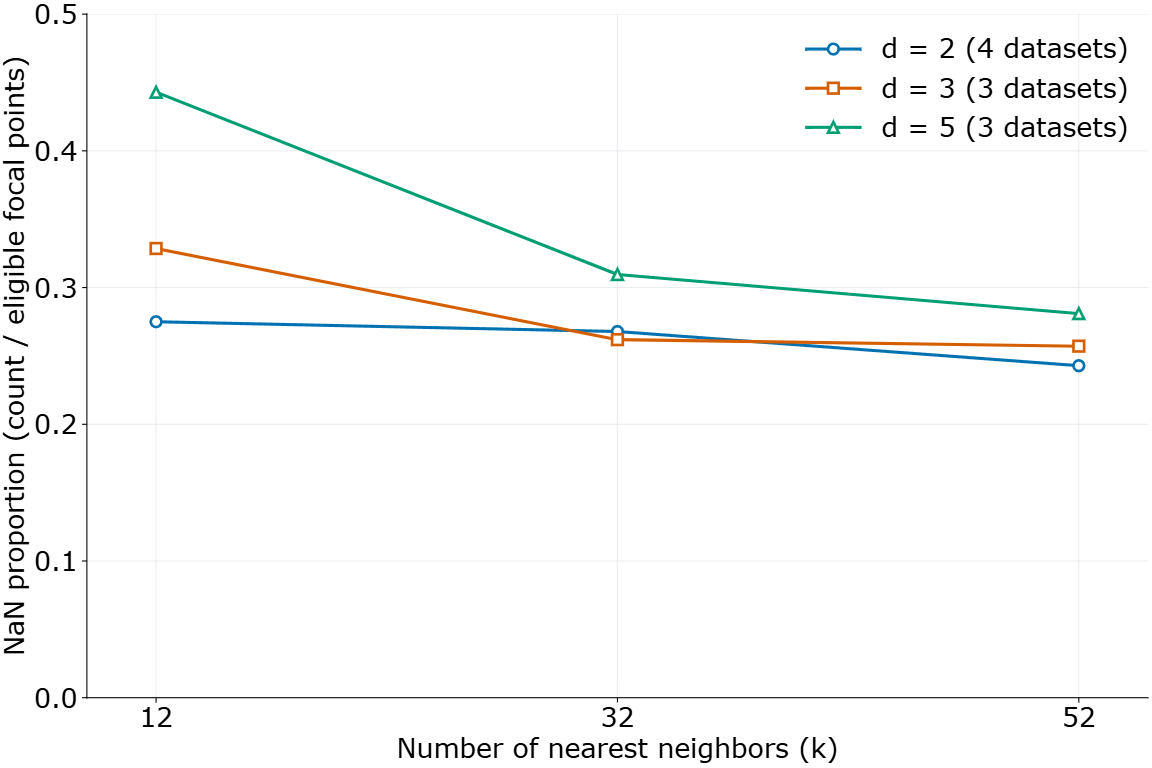}
    \caption{Proportion of missing scores from the oriented cone-based statistic in the local-sector evaluation setting.}
    \label{fig:oriented_nan_local}
\end{figure}

\begin{landscape}
\section{Boundary Estimates Benchmark}\label{annex:boundary_benchmark}
    \begin{figure}[H]
        \centering
        \includegraphics[
            width=0.95\linewidth,
            height=0.85\textheight,
            keepaspectratio
        ]{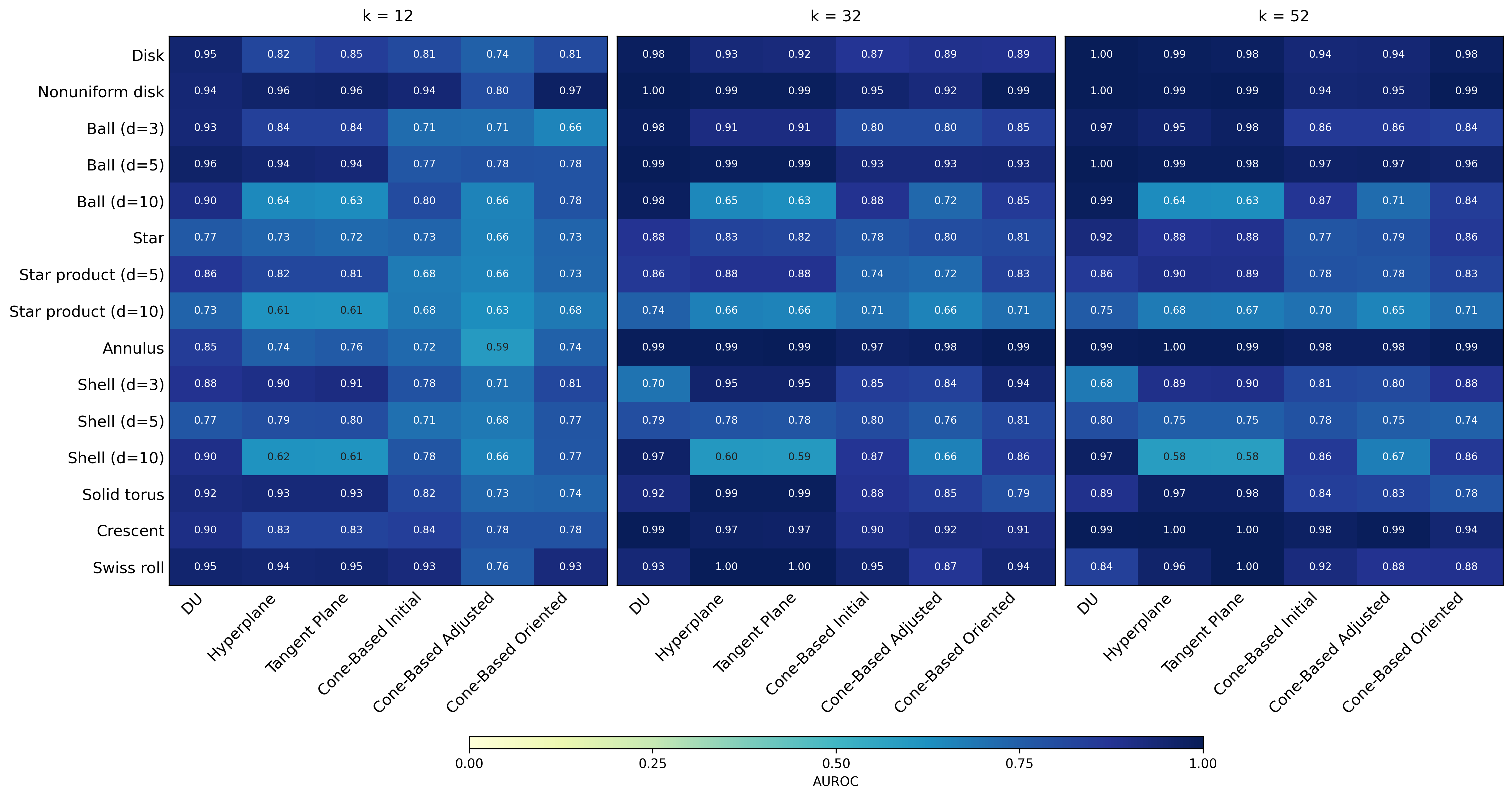}
        \caption{AUROC results of the SOTA and cone-based statistics for $k\in \{12,32,52\}$ in the support-boundary datasets.}
        \label{fig:support_boundary_cross}
    \end{figure}
    \begin{figure}[H]
        \centering
        \includegraphics[
            width=0.95\linewidth,
            height=0.85\textheight,
            keepaspectratio
        ]{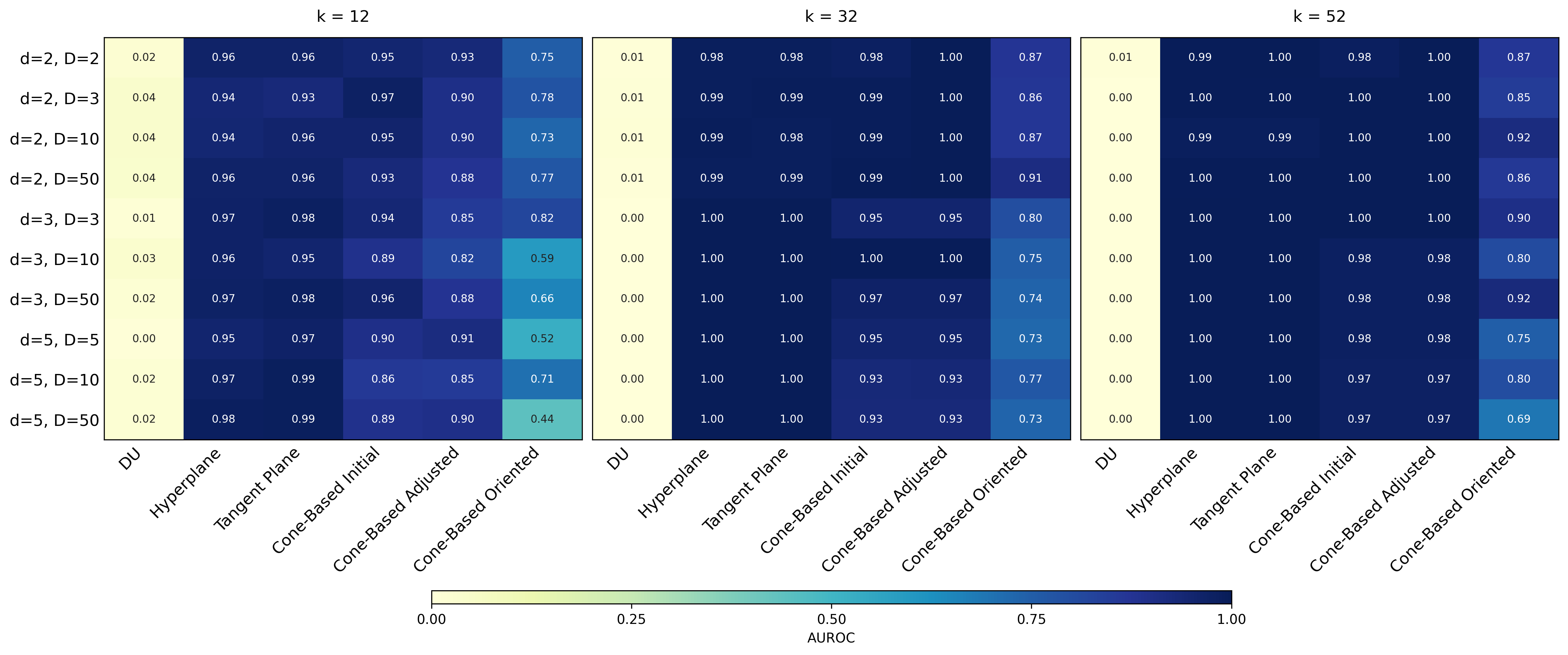}
        \caption{AUROC results of the SOTA and cone-based statistics for $k\in \{12,32,52\}$ in the local-sector datasets.}
        \label{fig:local_sector_cross}
    \end{figure}
    \begin{figure}[H]
        \centering
        \includegraphics[
            width=0.95\linewidth,
            height=0.85\textheight,
            keepaspectratio
        ]{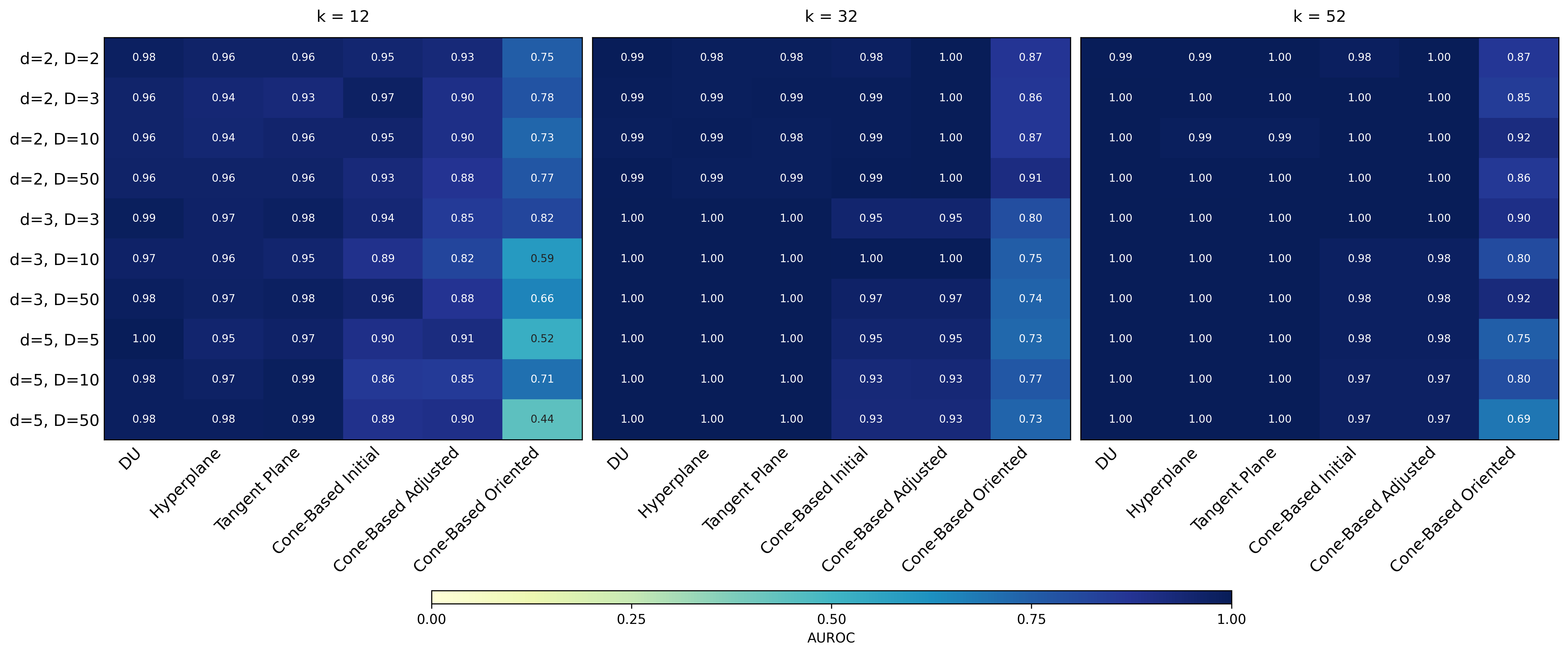}
        \caption{Inverted AUROC results of the DU estimates compared to original AUROC of the SOTA and cone-based statistics for $k\in \{12,32,52\}$ in the local-sector datasets.}
        \label{fig:inverted_DU}
    \end{figure}
\end{landscape}

\section{Application}\label{annex:application}

\begin{figure}[H]
    \centering
    \includegraphics[width=0.8\linewidth]{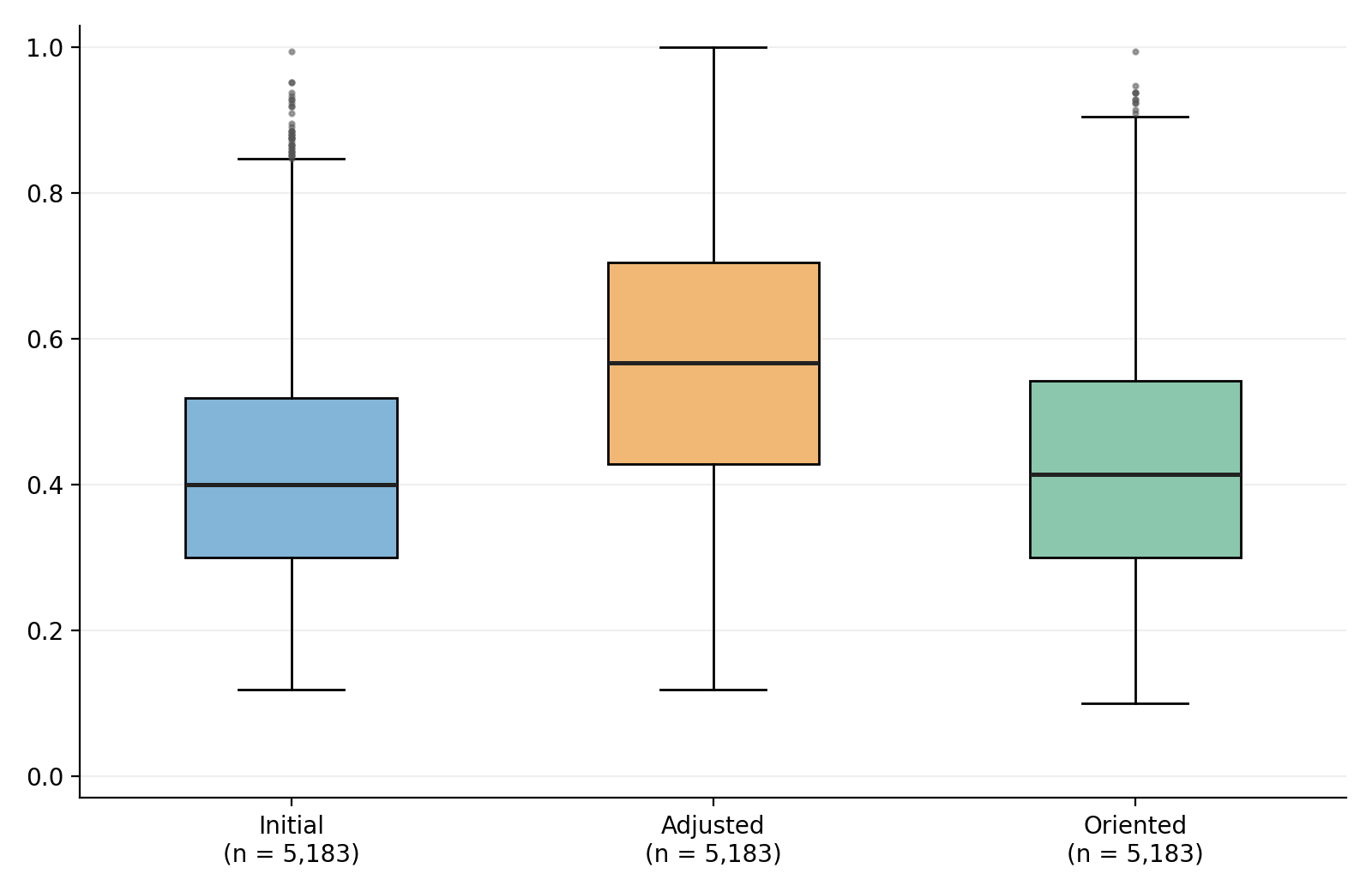}
    \caption{Proportion of boundary and interior  behaviors across $14$ PCs provided by each variant. A point with a high $P_i^{v}$ indicates that this point is frequently at the boundary of its neighborhood.}
        \label{fig:persistent_bos_variant}
\end{figure}
We notice that oriented boundary scores are missing for $572$ points in specific $D_B$ due to the alignment constraint. For these observations, the persistent oriented score is computed as the mean over available boundary scores across $D_B$.

 The PCA dimensions we consider for the RAG workflow are $D_{\text{Reduced}}=\{16,32,64,128,256\}$. Given the following constraint to collect non-zero PC directions,
 \begin{align*}
 D_{\text{Reduced}}\leq \min(1024,N_{B}-1),
 \end{align*}
 
 we set the persistent $P_i^{v}$ threshold, \textit{Pers\_THRESH}, in each variant such that $N_{B}-1\geq 256$, the maximal PC dimension of the set $d$. Setting Pers\_THRESH$=0.8$ for all three variants (Table \ref{tab:persistent_boundary_counts}) is too strict for the initial and oriented variants to meet the above constraint. Hence, we relax this threshold to $0.7$ only for these two.

\begin{table}[H]
\centering
\caption{Counts of points satisfying the selected persistent boundary
score thresholds.}
\label{tab:persistent_boundary_counts}
\begin{tabular}{lrrr}
\hline
Variant & $P_i\geq 0.8$ & $P_i\geq 0.7$ \\
\hline
Initial  & 80 & 289 \\
Adjusted & 588  & --  \\
Oriented & 104 & 365 \\
\hline
\end{tabular}
\end{table}

Using these three boundary sets, each matrix has a rank of at least $256$. We provide as well the overlap between boundary sets in the table below.

\begin{table}[H]
    \centering
    \caption{Number of overlapping boundary points between sets
    (initial and oriented thresholds: 0.7; adjusted threshold: 0.8).}
    \label{tab:boundary_set_overlap}
    \begin{tabular}{lr}
        \hline
        Boundary point sets & Shared points \\
        \hline
        Initial $\cap$ Adjusted  & 270 \\
        Initial $\cap$ Oriented  & 237 \\
        Adjusted $\cap$ Oriented & 293 \\
        \hline
    \end{tabular}
\end{table}

\end{document}